\documentclass{jpp}
\usepackage[latin9]{inputenc}
\usepackage{amsbsy}
\usepackage{graphicx}
\usepackage{esint}
\usepackage[unicode=true,
 bookmarks=false,
 breaklinks=false,pdfborder={0 0 1},backref=section,colorlinks=false]
 {hyperref}

\makeatletter
\usepackage{graphics}
\usepackage{epstopdf}
\usepackage{epsfig}

\newcommand{\apj}{ApJ}

\newcommand{\apjl}{ApJL}
\newcommand{\jgr}{J. Geophys. Res.}
\newcommand{\aap}{A~\&~A}
\newcommand{\aapr}{Astron.~\&~Astrophys.~Rev.}
\newcommand{\mnras}{MNRAS}
\newcommand{\solphys}{Sol.Phys.}
\newcommand{\physrep}{Phys. Rep.}

\newcommand{\ssr}{Space Sci. Rev.}

\shortauthor{V.V.Pipin}

\title{The magnetic helicity density patterns from non-axisymmetric solar
dynamo}

\author{Valery V. Pipin}
\affiliation{
Institute for Solar-Terrestrial Physics, 
PO Box 291, Lermontov st., 126a, 
Irkutsk, 664033, Russia}

\makeatother

\begin{document}

\title{The magnetic helicity density patterns from non-axisymmetric solar
dynamo}

\maketitle 
\begin{abstract}
{In the paper we study the helicity density patterns which can result
from the emerging bipolar regions. Using the relevant dynamo model
and the magnetic helicity conservation law we find that the helicity
density pattern around the bipolar regions depends on the configuration
of the ambient large-scale magnetic field, and in general they show
the quadrupole distribution. The position of this pattern relative
to the equator can depend on the tilt of the bipolar region. We compute
the time-latitude diagrams of the helicity density evolution. The
longitudinally averaged effect of the bipolar regions show two bands
of sign for the density distribution in each hemisphere. Similar helicity
density patterns are provided by the helicity density flux from the
emerging bipolar regions subjected to the surface differential rotation.
} 
\end{abstract}


\section{Introduction}

The magnetic helicity conservation is often considered as one of the
most important ingredient in the magnetic field generation on the
Sun and other solar type stars \citep{Brandenburg2005b,Blackman2015}
It is also important for the magnetic activities in the solar atmosphere
and corona \citep{Mackay2012}. The magnetic helicity characterize
the complexity of the magnetic field topology in the closed volume
\citep{Berger1984}. Observing the magnetic field on the surface we
can deduce some local proxies of the helicity integral. The observational
constraints of the solar dynamo models are related to the hemispheric
helicity rules \citep{Seehafer1994,Pevtsov1994} and the magnetic
helicity fluxes from the solar interior \citep{Berger2000,Blackman2003}.
The hemispheric helicity rule (hereafter HHR) follows from the theoretical
properties of the large-scale dynamo which we expect to be working
in the solar interior. This rule states that the small-scale magnetic
fields on the northern hemisphere have the negative twist (the right-hand
coordinate system) and the opposite twist is in the southern hemisphere.
In the dynamo theory the small scales include scale of the solar active
regions and smaller ones. The helicity conservation constrains the
helicity sign distribution over the spatial scales. The results of
\citet{Pouquet1975} showed that in the turbulent dynamo processes
we can expect that the twist of the large-scale magnetic field is
opposite to the twist of the small-scales magnetic field. This property
is also called the bi-helical dynamo.

The vector magnetic field observations make possible to deduce the
information about the magnetic and current helicity density on the
solar surface \citep{Pevtsov1994,Bao1998,Zhang2010}. In general,
theses proxies show the HHR for the small-scale magnetic field. The
issues about the HHR of the large-scale magnetic field and the bi-helical
solar dynamo are still debatable \citep{Brandenburg2017,Singh2018,Pipin2019a}.

However, the relevance of the local proxies, which are observed on
the surface, as the proxies of the bi-helical magnetic fields generated
by the dynamo processes in the depth of the convection zone can be
questioned. On the surface some of the helical magnetic field configurations
can be generated by means of other processes which are not readily
related to the dynamo. For example, the emerging of the magnetic field
on the surface and its interaction with the ambient magnetic field
can produce the local helicity flux. \citet{Hawkes2019AA} showed
one of such example, see Fig4c in their paper. They illustrated that
the emerging bipolar region which interacts with the global magnetic
field results to the local quadrupole helicity density flux pattern.
Interesting that this effect drives the helicity density flux butterfly
diagrams, which are satisfying the own HHR, see the Fig3c in their
paper. Another kind of this effect was illustrated by \citet{Pipin2019a}
in their benchmark dynamo model. Here we elaborate on this example
and study the effect of the emerging bipolar regions on the surface
helicity density patterns. Note, on the solar surface the most magnetic
activity is produced by the sunspots \citep{Stenflo2013a}. We find
that the emerging active regions can bias our conclusions about the
helical properties of the dynamo inside the convection zone. The study
is based on the numerical simulations of the nonaxisymmetric dynamo
model, which was suggested recently by \citet{Pipin2018d}. The next
section describes the model.

\section{Dynamo model}

Evolution of the large-scale magnetic field in perfectly conductive
media is described by the mean-field induction equation \citep{Krause1980}:
\begin{equation}
\partial_{t}\left\langle \mathbf{B}\right\rangle =\mathbf{\nabla}\times\left(\mathbf{\mathbf{\mathbf{\mathcal{E}}}+}\left\langle \mathbf{U}\right\rangle \times\left\langle \mathbf{B}\right\rangle \right)\label{eq:mfe}
\end{equation}
where $\mathbf{\mathcal{E}}=\left\langle \mathbf{u\times b}\right\rangle $
is the mean electromotive force; $\mathbf{u}$ and $\mathbf{b}$ are
the turbulent fluctuating velocity and magnetic field respectively;
and $\left\langle \mathbf{U}\right\rangle $ and $\left\langle \mathbf{B}\right\rangle $
are the mean velocity and magnetic field. \citet{Pipin2018d} (PK18)
suggested the minimal set of the dynamo equations to model the non-axisymmetric
magnetic field evolution. In this model, similarly to \citet{Moss2008},
we neglect the radial dependence of magnetic field, and assume that
the radial gradient of angular velocity is greater than the latitudinal
gradient. In earlier studies, e.g., \citet{Jennings1990} it was found
that neglecting of terms related to r-dependence may cause significant
reduction of the threshold for magnetic field excitation by a dynamo
process. Therefore, we construct the model in such a way that it holds
the results of \citet{Moss2008} in the limit of the axisymmetric
1D case. Therefore the results of the model have the illustrative
character which is typical for the toy dynamo models.

It is convenient to represent the vector $\left\langle \mathbf{B}\right\rangle $
in terms of the axisymmetric and non-axisymmetric components as follows:
\begin{eqnarray}
\left\langle \mathbf{B}\right\rangle  & = & \overline{\mathbf{B}}+\tilde{\mathbf{B}}\label{eq:b0}\\
\mathbf{\overline{B}} & = & \hat{\mathbf{\phi}}B+\nabla\times\left(A\hat{\mathbf{\phi}}\right)\label{eq:b1}\\
\tilde{\mathbf{B}} & = & \mathbf{\nabla}\times\left(\mathbf{r}T\right)+\mathbf{\nabla}\times\mathbf{\nabla}\times\left(\mathbf{r}S\right),\label{eq:b2}
\end{eqnarray}
where $\overline{\mathbf{B}}$ and $\tilde{\mathbf{B}}$ are the axisymmetric
and non-axisymmetric components; ${A}$, ${B}$, ${T}$ and ${S}$
are scalar functions representing the field components; $\hat{\mathbf{\phi}}$
is the azimuthal unit vector, $\mathbf{r}$ is the radius vector;
$r$ is the radial distance, and $\theta$ is the polar angle. {
\citet{Krause1980} showed that the only gauge transformation for
potentials T and S is a sum with the arbitrary r-dependent function.
To fix this arbitrariness they suggested the following normalization:}
\begin{equation}
\int_{0}^{2\pi}\int_{-1}^{1}Sd\mu d\phi=\int_{0}^{2\pi}\int_{-1}^{1}Td\mu d\phi=0.\label{eq:norm-1}
\end{equation}

Hereafter, the over-bar denotes the axisymmetric magnetic field, and
tilde denotes non-axisymmetric properties. Following the above ideas
we consider a reduced dynamo model where we neglect the radial dependence
of the magnetic field. In this case, the induction vector of the large-scale
magnetic field is represented in terms of the scalar functions as
follows: 
\begin{eqnarray}
\left\langle \mathbf{B}\right\rangle  & = & -\frac{\mathbf{\hat{r}}}{R^{2}}\frac{\partial\sin\theta A}{\partial\mu}-\frac{\hat{\theta}}{R}A+\hat{\mathbf{\phi}}B\label{eq:3d}\\
 & - & \frac{\hat{\mathbf{r}}}{R^{2}}\Delta_{\Omega}S+\frac{\hat{\theta}}{\sin\theta}\frac{\partial T}{\partial\phi}+\hat{\phi}\sin\theta\frac{\partial T}{\partial\mu},\nonumber 
\end{eqnarray}
where $R$ represents the radius of the spherical surface inside a
star where the hydromagnetic dynamo operates. The above equation defines
the 3d divergency free B-field on the sphere. In the model we employ
the simple expression of $\mathbf{\mathcal{E}}$:

\begin{eqnarray}
\mathbf{\mathcal{E}} & = & \alpha\left\langle \mathbf{B}\right\rangle -\eta_{T}\mathbf{\nabla}\times\mathbf{\left\langle B\right\rangle }+V_{\beta}\hat{\mathbf{r}}\times\left\langle \mathbf{B}\right\rangle +\alpha_{\beta}\hat{\mathbf{\phi}}\left\langle \mathbf{B}\right\rangle _{\phi}\label{eq:simpE}
\end{eqnarray}
The last term is introduced to simulated the tilt of the emerging
active regions. The magnetic buoyancy is source of the nonaxisymmetric
magnetic field in the model. We assume that the magnetic buoyancy
acts on relatively small-scale parts of the axisymmetric magnetic
field, perhaps, it is caused by some kind of nonlinear instability.
 {This effect can not be derived within the mean-field magneto-hydrodynamic
framework. The main purpose for introducing this effect is to mimic
the bipolar region formation on the solar surface. Following to results
of \citet{Pipin2018d}, the effect satisfactory reproduces the bipolar
structure of the radial magnetic field which can observed in the simple
active regions. We assume that the magnetic buoyancy velocity depends
on the strength of the large-scale magnetic field and the random longitudinal
position.} It is formulated as following: 

\begin{eqnarray}
 &  & \frac{\mathrm{\alpha_{MLT}}}{\gamma}u_{c}\beta^{2}K(\beta)(1+\xi_{\beta}(\phi,\theta)),\beta\ge\beta_{cr}\nonumber \\
V_{\beta}=\label{eq:buoy}\\
 &  & 0,\beta<\beta_{cr},\nonumber 
\end{eqnarray}
where, $\mathrm{\alpha_{MLT}}=1.9$ is the mixing-length theory parameter,
$\gamma$ is the adiabatic law constant, $\beta=\left|\left\langle \mathbf{B}\right\rangle \right|/\mathrm{B_{eq}}$,
$\mathrm{B_{eq}}=\sqrt{4\pi\overline{\rho}u_{c}^{2}}$, function $K\left(\beta\right)$
is defined in \citet{Kitchatinov1993}. For $\beta\ll$1, we have
$K\left(\beta\right)\sim1$, and for the strong magnetic field, when
$\beta>$1, { $K\left(\beta\right)\sim1/\beta^{3}$. }The function
$\xi_{\beta}\left(\phi,\theta\right)$ describes the latitudinal and
longitudinal dependence of the instability, and parameter $\beta_{cr}$
controls the instability threshold. In this formulation, the preferable
latitude of the ``active region emergence'' is determined by the
latitude of the maximum of the toroidal magnetic field energy, $\theta_{\mathrm{max}}$.
The magnetic buoyancy instability perturbations are determined by
function: 
\begin{equation}
\xi_{\beta}\left(\phi,\theta\right)=C_{\beta}\exp\left(-m_{\beta}\left(\sin^{2}\left(\frac{\phi-\phi_{0}}{2}\right)+\sin^{2}\left(\frac{\theta-\theta_{\mathrm{max}}}{2}\right)\right)\right),\label{eq:xib}
\end{equation}
The instability is randomly initiated in the northern or southern
hemispheres, and the longitude, $\phi_{0}$, is chosen randomly, as
well. We arbitrary chose the fluctuation interval $\tau_{\beta}=0.01P_{cyc}$,
where $P_{cyc}$ is the dynamo cycle period. The dynamic of the buoyancy
instability is restricted by the five time run-steps, i.e., we put
the function $\xi_{\beta}\left(\phi,\theta\right)$ to zero after
five steps from initiation. The given period roughly corresponds to
a few days. In the model we measure the time in the diffusive units,
$R^{2}/\eta_{T}$. If we scale the dynamo period of the model to 11
years, then the simulated emerging time is about one week, which is
much longer than on the Sun \citep{Toriumi2019}. Parameter $m_{\beta}$
controls the spatial scale of the instability. Theoretically, using
a high values of $m$ we can reproduce the spatial scale of the solar
active regions. However, this requires increasing the resolution both
in longitude and latitude. This make a toy model computationally expensive.
We chose the value $m_{\beta}=1/\delta\phi^{2}$, $\delta\phi=2^{\circ}$,
which allows us to accurately resolve the evolving nonaxisymmetric
perturbations of magnetic field, and qualitatively reproduce the essential
physical effects. Parameter $C_{\beta}$ controls the amount of the
injected magnetic flux. If large-scale toroidal magnetic flux at a
given co-latitude $\theta$ is transformed into magnetic flux of the
perturbation then $\left\langle \xi_{\beta}\left(\phi,t\right)\right\rangle _{\phi}\approx1$.
This condition corresponds to $C_{\beta}\approx15$. In reality, solar
active regions are formed by concentration of the toroidal magnetic
flux emerging in the photosphere. Turbulent convective motions and
other physical processes may take part in the process of formation
of solar active regions. Therefore, parameter $C_{\beta}$ can be
higher than the above mentioned value. Similarly to \citet{Pipin2018d},
we choose $C_{\beta}=40$. In the numerical experiments, we found
that higher values of $C_{\beta}$ result in strong cycle-to-cycle
variability of the magnetic energy even in the case of stationary
$\alpha$-effect. For the chosen $C_{\beta}$, $\left|\left\langle \mathbf{B}\right\rangle \right|\approx2\left|\overline{B_{\phi}}\right|$,
therefore fluctuations of the magnetic field because of the magnetic
buoyancy instability are of the order of the axisymmetric toroidal
magnetic field strength, which is widely accepted in the literature
\citep{Krause1980,Brandenburg2005b}. 

In the model the strength of the magnetic field is measured relative
to the strength of the equipartition field, $\mathrm{B_{eq}}$. Dependence
of the instability on the toroidal magnetic field strength results
in the magnetic cycle modulation in the number of the bipolar regions.
Besides, the critical parameter $\beta_{cr}=0.5$ prevents the emergence
of active regions at high latitudes.  {The $\beta_{cr}\le$1
ensures that the magnetic field strength in the emerged bipolar regions
is about the equipartition value.} To get the model closer to observations
the temporal and spatial resolution have to be increased. Bearing
in mind the whole simplicity of our model we restrict ourselves to
the qualitative considerations. In our description, the magnetic buoyancy
instability results in generation of the nonaxisymmetric magnetic
field in form of the magnetic bipolar regions. Besides, it results
in the magnetic flux loss and the large-scale dynamo saturation for
the magnetic field strength $\left|\boldsymbol{B}\right|>0.5\mathrm{B_{eq}}$
. This was anticipated earlier, e.g., by \citet{Parker1984} and \citet{Noyes1984}. 

In this paper we would like to calculate the magnetic helicity flux
provided by emergence of the tilted active regions on the solar surface.
For this purpose, we assume that the emerging part of the toroidal
magnetic field is subjected to some extra $\alpha$ effect, which
is caused by the dynamic of the magnetic loop. Note, that the dynamic
of the buoyancy instability is restricted to 5 time run-steps. With
this choice, we find that a 5$^{\circ}$ tilt \citep{Tlatov2013}
can be reproduced if we put 
\begin{equation}
\alpha_{\beta}=0.33\cos\theta V_{\beta},\label{eq:ab}
\end{equation}
where $V_{\beta}$ is determined by the Eq(\ref{eq:buoy}) and the
coefficient 0.33 was found by the numerical experiments. In case of
$\alpha_{\beta}=V_{\beta}$ the tilt is around $\pi/4$. Note, that
the first term in the equation (\ref{eq:simpE}) includes the $\alpha$-effect
which acts on the unstable part of the nonaxisymmetric magnetic field
and produces some amount of tilt as well. We find this effect negligible
in comparison observations. We deliberately assume that the $\alpha_{\beta}$
-effect acts only on the nonaxisymmetric part of the magnetic field.
In this case, it does not change the conditions for the linear stability
of the axisymmetric magnetic field. 

For the standard part of $\alpha$ effect we use the phenomenological
description and take into account the contribution of the magnetic
helicity in following to suggestions of \citet{Pouquet1975}: 
\begin{eqnarray}
\alpha & = & \alpha_{0}\cos\theta+\frac{\left\langle \mathbf{b}\cdot\nabla\times\mathbf{b}\right\rangle \tau_{c}}{4\pi\overline{\rho}}\label{alp2d}
\end{eqnarray}
where $\alpha_{0}$ is a free positive parameter which controls the
strength of the $\alpha$- effect. The results of the above-cited
paper suggest that the negative sign of the small-scale current helicity
results from the magnetic helicity conservation. In following to \citet{Moffatt1978}
we replace $\left\langle \mathbf{b}\cdot\nabla\times\mathbf{b}\right\rangle \sim\left\langle \chi\right\rangle /\ell^{2}$
, where $\left\langle \chi\right\rangle $ is the magnetic helicity
density, $\left\langle \chi\right\rangle =\left\langle \mathbf{a}\cdot\mathbf{b}\right\rangle $
($\mathbf{a}$ and $\mathbf{b}$ are the fluctuating parts of magnetic
field vector-potential and magnetic field vector). The magnetic helicity
conservation results to the dynamical quenching of the dynamo \citep{Kleeorin1982,Kleeorin1999,Kleeorin2000}.
For the sake of simplicity the above equation (\ref{eq:ab}) has no
explicit inclusion of magnetic helicity. It is partly justified by
the fact that in our formulation the $\alpha_{\beta}$ term is not
a primary dynamo effect. This is in contrast with the Babcock-Leighton
scenario for the solar dynamo \citep{Charbonneau2011}. In following
\citep{Blackman2003}, we can anticipate that explicit addition of
the magnetic helicity term in the equation (\ref{eq:ab}) will result
in reduction of the bipolar region tilt. Also, in our we do not include
any algebraic quenching (see, \citealp{Ruediger1993b}) in the $\alpha$
-effect contributions in Eq(\ref{eq:ab}) and Eq(\ref{alp2d}). For
a simple plane model like ours the only result of the algebraic $\alpha$-quenching
is the reduction of the magnitude of the generated toroidal magnetic
field. We find that the model reproduce the finite amplitude dynamo
waves of the reference model of \citet{Moss2008} even for the case
when we have the only nonlinear effect due to the magnetic buoyancy
term. Therefore, we skip the algebraic $\alpha$-quenching in our
calculations.

Similarly to \citet{Hubbard2012,Pipin2013c,Brandenburg2018}, in our
model we use the global conservation law for the total magnetic helicity.
In this case the magnetic helicity density, $\left\langle \chi\right\rangle =\left\langle \mathbf{a}\cdot\mathbf{b}\right\rangle $,
is governed by the equation: 
\begin{equation}
\left(\frac{\partial}{\partial t}+\boldsymbol{\left\langle \mathbf{U}\right\rangle \cdot\nabla}\right)\left\langle \chi\right\rangle ^{(tot)}=-\frac{\left\langle \chi\right\rangle }{R_{m}\tau_{c}}-2\eta\left\langle \mathbf{B}\right\rangle \cdot\left\langle \mathbf{J}\right\rangle -\mathbf{\nabla\cdot}\mathbf{\mathbf{\mathcal{F}}}^{\chi},\label{eq:helcon}
\end{equation}
where $\left\langle \chi\right\rangle ^{(tot)}=\left\langle \chi\right\rangle +\left\langle \mathbf{A}\right\rangle \cdot\left\langle \mathbf{B}\right\rangle $
is the total magnetic helicity density of the mean and turbulent fields.
It is assumed that $\nabla\cdot\left\langle \mathbf{U}\right\rangle =0$.
Note, that in derivations of the Eq.(\ref{eq:helcon}), in following
\citet{Kleeorin1999}, we have ${\displaystyle 2\eta\mathbf{\left\langle b\cdot j\right\rangle }=\frac{\left\langle \chi\right\rangle }{R_{m}\tau_{c}}}$.
 {The heuristic term,} $\mathbf{\mathbf{\mathcal{F}}}^{\chi}=-\eta_{\chi}\mathbf{\nabla}\left\langle \chi\right\rangle ^{(tot)}$
is the diffusive flux of the total magnetic helicity, and $R_{m}$
is the magnetic Reynolds number. The coefficient of the turbulent
helicity diffusivity, $\eta_{\chi}$, is chosen ten times smaller
than the isotropic part of the magnetic diffusivity \citep{Mitra2010}:
$\eta_{\chi}=\frac{1}{10}\eta_{T}$. Here, in comparison to the axisymmetric
model of \citet{Pipin2013c}, we have to take into account the redistribution
of the nonaxisymmetric part magnetic helicity by the differential
rotation. Our ansatz differs from that suggested by papers of \citet{Kleeorin1982,Kleeorin1999}.
Here, the turbulent fluxes of the magnetic helicity are approximated
by the only term which is related to the diffusive flux. \citet{Pipin2013c}
found that the dynamo models, where helicity evolution follows the
Eq(\ref{eq:helcon}), show a magnetic helicity wave propagating with
the dynamo wave. This alleviates the so-called catastrophic quenching
of the $\alpha$-effect \citep{Brandenburg2018}.

Similarly to the magnetic field, the mean magnetic helicity density
can be formally decomposed into the axisymmetric and nonaxisymmetric
parts: $\left\langle \chi\right\rangle ^{(tot)}=\overline{\chi}^{(tot)}+\tilde{\chi}^{(tot)}$.
The same can be done for the magnetic helicity density of the turbulent
field: $\left\langle \chi\right\rangle =\overline{\chi}+\tilde{\chi}$,
where $\overline{\chi}=\overline{\mathbf{a}\cdot\mathbf{b}}$ and
$\tilde{\chi}=\tilde{\left\langle \mathbf{a}\cdot\mathbf{b}\right\rangle }$.
Then we have, 
\begin{eqnarray}
\overline{\chi}^{(tot)} & = & \overline{\chi}+\overline{\mathbf{A}}\cdot\overline{\mathbf{B}}+\overline{\tilde{\mathbf{A}}\cdot\tilde{\mathbf{B}}},\label{eq:t1}\\
\tilde{\chi}^{(tot)} & = & \tilde{\chi}+\overline{\mathbf{A}}\cdot\tilde{\mathbf{B}}+\tilde{\mathbf{A}}\cdot\overline{\mathbf{B}}+\tilde{\mathbf{A}}\cdot\tilde{\mathbf{B}},\label{eq:t2}
\end{eqnarray}
Evolution of the $\overline{\chi}$ and $\tilde{\chi}$ is governed
by the corresponding parts of Eq(\ref{eq:helcon}). Thus, the model
takes into account contributions of the axisymmetric and nonaxisymmetric
magnetic fields in the whole magnetic helicity density balance, providing
a non-linear coupling. We see that the $\alpha$-effect is dynamically
linked to the longitudinally averaged magnetic helicity of the $\tilde{\mathbf{B}}$-field,
which is the last term in Eq(\ref{eq:t1}). Thus, the nonlinear $\alpha$-effect
is\emph{ nonaxisymmetric}, and it contributes into coupling between
the $\overline{\mathbf{B}}$ and $\tilde{\mathbf{B}}$ modes. The
coupling works in both directions. For instance, the azimuthal $\alpha$-effect
results in $\mathcal{E}_{\phi}=\alpha\left\langle B_{\phi}\right\rangle +\alpha_{\beta}\tilde{B}_{\phi}$.
If we denote the nonaxisymmetric part of the $\alpha$ by $\tilde{\alpha}$
then the mean electromotive force is $\overline{\mathcal{E}}_{\phi}=\overline{\alpha}\overline{B}_{\phi}+\overline{\tilde{\alpha}\tilde{B}_{\phi}}+\overline{\alpha_{\beta}\left\langle B\right\rangle _{\phi}}$.
This introduces a new generation source which is usually ignored in
the axisymmetric dynamo models. The magnetic helicity conservation
is determined by the magnetic Reynolds number $R_{m}$. In this paper
we employ $R_{m}=10^{6}$.

The helicity conservation in form the Eq.(\ref{eq:helcon}) is suitable
for the dynamo simulation. To estimate the helicity flux from the
dynamo we will follow the approach of \citet{Berger2000}. We derive
the equation for the small-scale helicity integral in Appendix\ref{helint}.
Following their consideration the change of the helicity integral
is determined by the the dynamo processes inside and the helicity
fluxes out of the dynamo regions as follows 

\begin{eqnarray}
\left(\frac{d}{dt}+\frac{1}{R_{m}\tau_{c}}\right)\int\left\langle \chi\right\rangle dV & = & -2\int\mathbf{\mathbf{\mathbf{\mathcal{E}}}}\cdot\left\langle \mathbf{B}\right\rangle \mathrm{dV}-\int\boldsymbol{\boldsymbol{\mathcal{F}}}^{\chi}\cdot\mathbf{n}\mathrm{dS}\label{helbr}\\
 &  & -2\oint\left(\left\langle \mathbf{A}\right\rangle \cdot\left\langle \mathbf{U}\right\rangle \right)\left(\left\langle \mathbf{B}\right\rangle \cdot\mathbf{n}\right)\mathrm{dS}-2\oint\left(\mathbf{\mathbf{\mathcal{E}}}\times\mathbf{\left\langle A\right\rangle }\right)\cdot\mathbf{n}\mathrm{dS}\nonumber \\
 &  & -2\eta\oint\left(\left\langle \mathbf{A}\right\rangle \times\left\langle \mathbf{J}\right\rangle \right)\cdot\mathbf{n}\mathrm{dS}+\oint\left(\left\langle \mathbf{A}\right\rangle \cdot\left\langle \mathbf{B}\right\rangle \right)\left(\left\langle \mathbf{U}\right\rangle \cdot\mathbf{n}\right)\mathrm{dS}\nonumber 
\end{eqnarray}
This equation is fully compatible with the eq\ref{eq:helcon}. It
is suitable for estimation of the helicity fluxes out of the dynamo
domain. The first term in the second line of the equation (\ref{helbr})
define the helicity flux due to the differential rotation:

\begin{equation}
F_{\Omega}=-2\left\langle B_{r}\right\rangle \left\langle A_{\phi}\right\rangle \overline{U}_{\phi}.\label{fldr}
\end{equation}
The second term, i.e., $-2\oint\left(\mathbf{\mathbf{\mathcal{E}}}\times\mathbf{\left\langle A\right\rangle }\right)\cdot\mathbf{n}\mathrm{dS}$
describes the helicity fluxes due effect of the turbulent flows and
magnetic field. The expression of the mean electromotive force contains
contributions from the $\alpha$ -effect, turbulent diffusivity and
the magnetic buoyancy. For this study we skip the effect of the turbulent
diffusivity. Note that this flux is additive to the term in the third
line of the Eq(\ref{helbr}). In our discussion, we leave contributions
due to the $\alpha$ and magnetic buoyancy effects:

\begin{eqnarray}
-2\left(\mathbf{\mathbf{\mathcal{E}}}\times\mathbf{\left\langle A\right\rangle }\right)\cdot\mathbf{n} & = & F_{\alpha}+F_{\beta}+\dots,\label{eq:eafl}\\
F_{\alpha} & = & -2\alpha\left(\left\langle B_{\theta}\right\rangle \left\langle A_{\phi}\right\rangle -\left\langle B_{\phi}\right\rangle \left\langle A_{\theta}\right\rangle \right)+2\alpha_{\beta}\left\langle B_{\phi}\right\rangle \left\langle A_{\theta}\right\rangle \\
F_{\beta} & = & 2V_{\beta}\left(\left\langle B_{\theta}\right\rangle \left\langle A_{\theta}\right\rangle +\left\langle B_{\phi}\right\rangle \left\langle A_{\phi}\right\rangle \right)
\end{eqnarray}
Note, that both the axisymmetric and the nonaxisymmetric modes contribute
to the all terms in Eq(\ref{eq:eafl}). 

The equations (\ref{eq:mfe},\ref{eq:helcon}) are solved numerically
in the non-dimensional form. We assume that the rotational shear is
constant in latitude. The effect of differential rotation is controlled
by non-dimensional parameter $R_{\omega}={\displaystyle \frac{R^{2}}{\eta_{T}}\frac{\partial\Omega}{\partial r}}$,
the $\alpha$-effect is measured by parameter $R_{\alpha}={\displaystyle \frac{R\alpha_{0}}{\eta_{T}}}$,
the magnetic buoyancy depends on $R_{\beta}={\displaystyle \frac{R}{\eta_{T}}\frac{\mathrm{\alpha_{MLT}}u_{c}}{\gamma}}$,
and the magnetic field is measured relative to the equipartition strength
$\mathrm{B_{eq}}=\sqrt{4\pi\overline{\rho}u'^{2}}$. Similarly to
\citet{Pipin2018d} we put $R_{\omega}={\displaystyle \frac{R\Omega}{\eta_{T}}=}10^{3}$,
$R_{\alpha}=1$. This choice describes the $\alpha^{2}\Omega$ dynamo
regime with differential rotation as the main driver of axisymmetric
toroidal magnetic field. Note for the given choice of the dynamo parameters
the nonaxisymmetric modes are stable. They do not take part in the
dynamo unless some nonaxisymmetric phenomena come into the play. In
this model the nonaxisymmetric modes are resulted due to the magnetic
buoyancy effect. The $R_{\beta}$ controls both the magnetic field
strength in bipolar regions and magnetic flux loss. Therefore, it
affects the level of the large-scale magnetic field strength in the
stationary state. To estimate the magnetic buoyancy parameter we employ
results of \citep{Kitchatinov1993} who argued that the maximum buoyancy
velocity of large-scale magnetic field of equipartition strength $B_{eq}$
is of the order of 6 m/s. In the solar conditions, the magnetic diffusion
$\eta_{T}=10^{12}$cm$^{2}/$s \citep{Ruediger2011}, and $R_{\beta}\approx500$. {
}With this value we get a very efficient magnetic flux loss, which
results in the large-scale toroidal magnetic field strength much less
than $B_{eq}$ . This regime is not efficient for the bipolar region
production. Hence, we use by an order of magnitude smaller value:
$R_{\beta}=50$. For this value, we get the ``spot's'' magnetic
field strength around $B_{eq}$. Reduction of the $R_{\beta}$ results
in the weaker bipolar regions and the stronger larger-scale toroidal
magnetic field. Note, that the magnetic helicity in the model is measured
in units $\mathrm{B_{eq}^{2}R}$. Comparing our results with observations,
we have to bear in mind that in the model the magnetic field dynamo
generation and the bipolar region formation occur in the same place.
Therefore, the resulted configuration of the axisymmetric magnetic
field are expected to be different from the solar surface observation.
However, the evolution of the nonaxisymmetric magnetic field mimics
the observational magnetic patterns reasonably well (see, \citealp{Pipin2018d}).
The further detail about the model can be found in the above cited
paper. Also, the python code for the model can be found at zenodo
: \citet{Pipin2018}.

\section{Results}

\subsection{Helicity density patterns from bipolar regions}

In this section we consider the helicity patterns, which are produced
by the emerging bipolar regions. In this case, we start simulations
with a simple antisymmetric distribution of the toroidal magnetic
field, $\overline{B}_{\phi}=2\sin2\theta$. The two bipolar regions
are injected successively in the southern and northern hemispheres
with interval about 0.004$R^{2}/\eta_{T}$. The time in the model
is measured in the units of the diffusive time. If we scale the dynamo
period of the model to 11 years, then the interval 0.004$R^{2}/\eta_{T}$
corresponds to 2.5 months.

\begin{figure}
\includegraphics[width=0.9\textwidth]{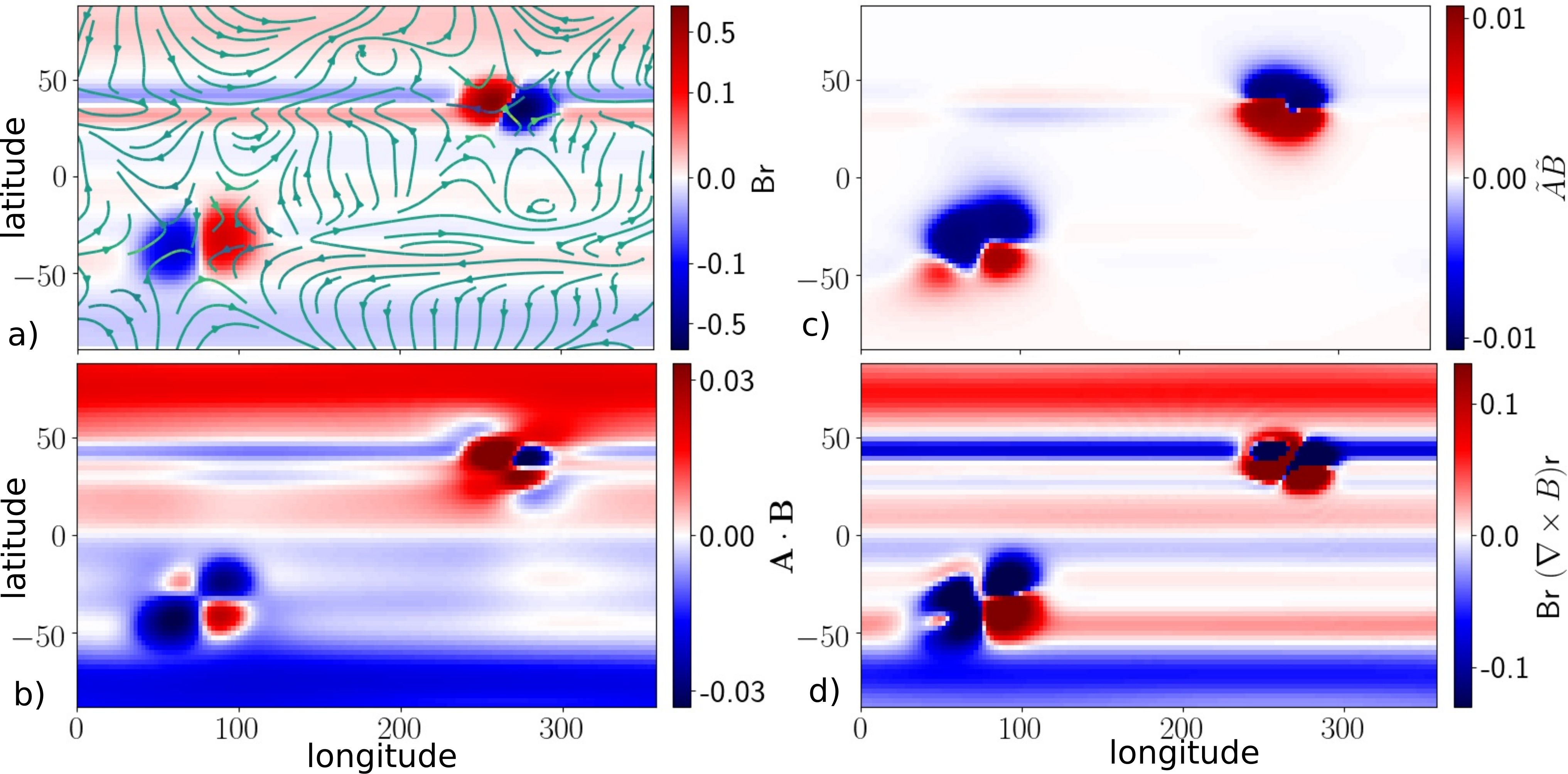}

\caption{\label{fig:bip}a) The color background shows the radial magnetic
field (both the axisymmetric and the nonaxisymmetric modes), streamlines
show the horizontal non-axisymmetric magnetic field; b) the total
magnetic helicity density, $\left(\overline{\mathbf{A}}+\tilde{\mathbf{A}}\right)\cdot\left(\overline{\mathbf{B}}+\tilde{\mathbf{B}}\right)$,
c) the same as b) for $\tilde{\mathbf{A}}\cdot\tilde{\mathbf{B}}$;
d) the same as b) for the current helicity density. }
\end{figure}

The Figure \ref{fig:bip} illustrates the magnetic field configurations,
as well as, the total helicity density, $\overline{\mathbf{A}}\cdot\overline{\mathbf{B}}+\tilde{\mathbf{A}}\cdot\tilde{\mathbf{B}}$,
the helicity density of the nonaxisymmetric magnetic fields and the
current helicity density distributions, $B_{r}\left(\nabla\times\mathbf{B}\right)_{r}$.
The snapshots are taken shortly after formation of the second bipolar
region in the northern hemisphere. We see that the helicity density
patterns of the bipolar regions have the quadrupole distributions.
The large-scale helicity density is in the background. In agreement
with the theoretical expectations, the large-scale magnetic field
has the positive magnetic and current helicity density sign in the
northern hemisphere. This helicity was generated by the large-scale
dynamo. The emerging bipolar regions show the inverted quadrupole
helicity patterns in the southern and the northern hemispheres. The
positive and negative helicity density parts nearly cancel each other
in each hemisphere. The effect of tilt most pronounced in distribution
of the current helicity density. On the $B_{r}\left(\nabla\times\mathbf{B}\right)_{r}$
synoptic map we see the negative trace, which is produced by the emerging
region. Therefore, $\overline{\tilde{B}_{r}\left(\nabla\times\mathbf{\tilde{B}}\right)_{r}}<0$
inside the latitudinal band of the bipolar region. We did not find
this effect in another run where we neglect the $\alpha_{\beta}$
term in the mean electromotive force. Also, some part of the net helicity
is due to participation of the bipolar regions in the large-scale
dynamo.

\begin{figure}
\includegraphics[width=0.9\textwidth]{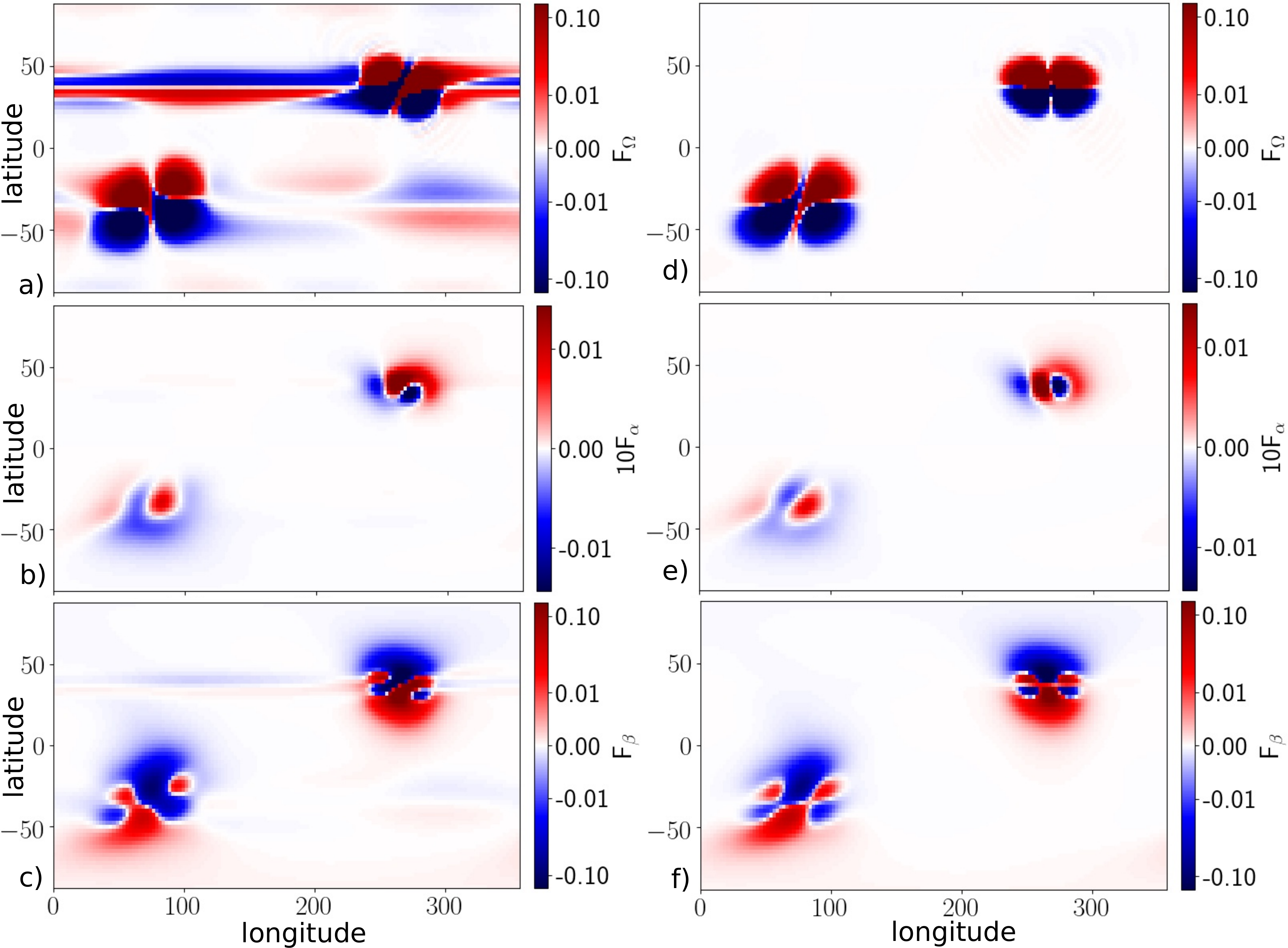}

\caption{\label{fig:flx}The helicity density fluxes, a) $F_{\Omega}$ - the
flux due to differential rotation; b) the flux due the $\alpha$ -
effect, $F_{\alpha}$; c) the flux due to the magnetic buoyancy, $F_{\beta}$;
the panels d),e) and f) shows the same for the model without tilt,
i.e., $\alpha_{\beta}=0$.}
\end{figure}

The Figure \ref{fig:flx} illustrates the patterns of the helicity
density fluxes, the terms, $F_{\Omega}$, $F_{\alpha}$ and $F_{\beta}$
in the Eq(\ref{helbr}) for the models with and without tilt effect.
We find that coupling the emerging bipolar regions and the differential
rotation produces the flux pattern which is inverted to the helicity
density, $\tilde{\mathbf{A}}\cdot\tilde{\mathbf{B}}$. The Figures
\ref{fig:flx}a) and d) agrees qualitatively with results of \citet{Hawkes2019AA}
(see, Fig.4c, there). It is found that the flux $F_{\beta}$ is substantially
smaller than the $F_{\Omega}$. Also, in all the cases the net helicity
flux from each bipolar region is close to zero.

\subsection{Bipolar regions in the dynamo evolution.}

We make a run of the dynamo model with the random injections of the
bipolar regions by means of the magnetic buoyancy instability. The
instability is randomly initiated in the northern or southern hemispheres,
and the longitude, $\phi_{0}$, is also chosen randomly. We arbitrary
chose the fluctuation interval $\tau_{\beta}=0.01P$. After injection
of the perturbation the evolution is solely determined by the dynamo
equations. Note, that the condition of the buoyancy instability is
defined by the critical magnetic field strength, see the Eq.(\ref{eq:buoy}).

\begin{figure}
\includegraphics[width=0.95\textwidth]{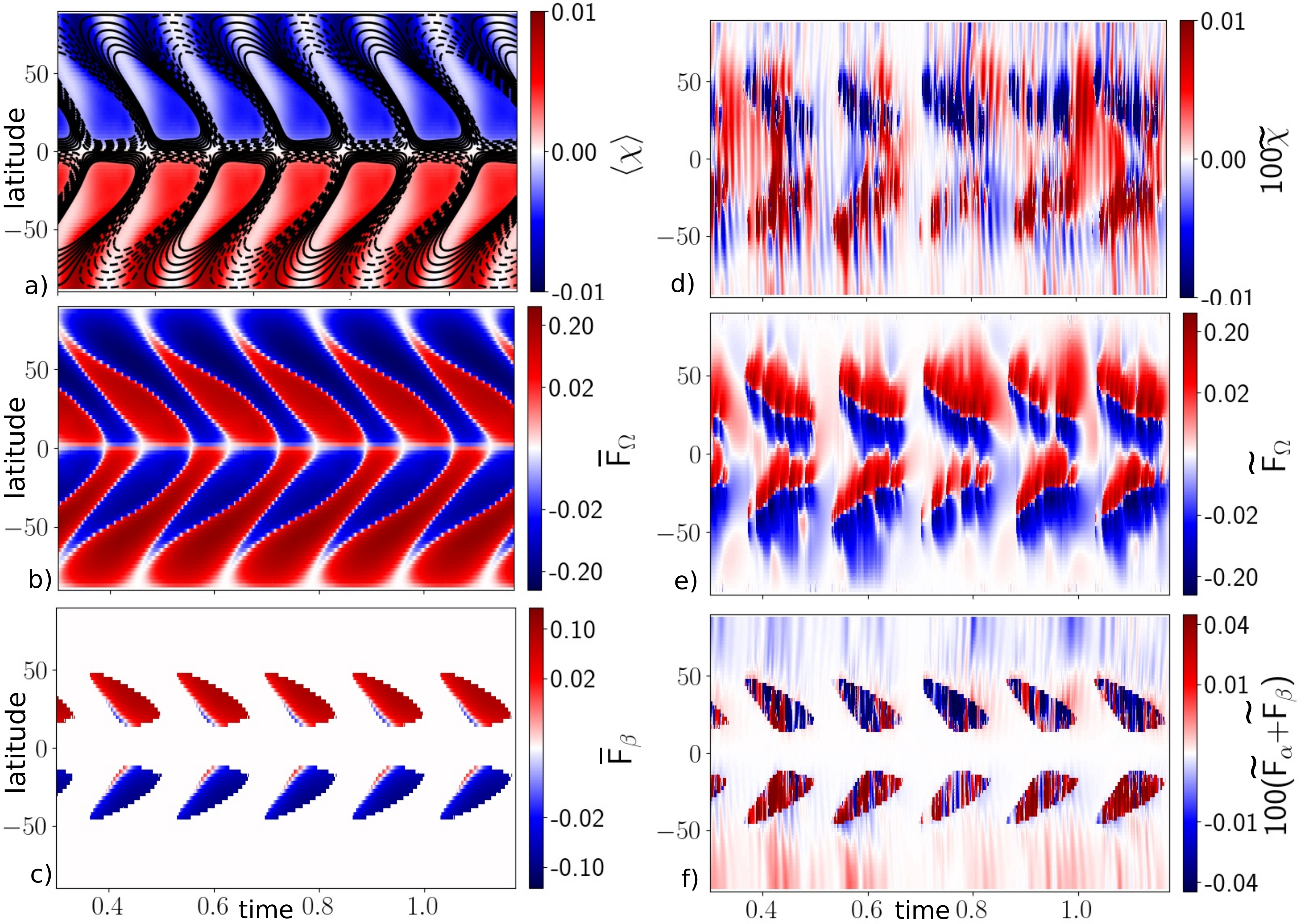}

\caption{\label{fig:tml}a)The time-latitude diagrams for the toroidal magnetic
field (contours) and the small-scale helicity density, $\left\langle \chi\right\rangle $,
is shown by the background image; b) the large-scale magnetic field
helicity flux due to the differential rotation; c) the same as b)
for the flux due to the magnetic buoyancy; d) the time-latitude evolution
of the helicity density of the nonaxisymmetric magnetic field; e)
show the same as b) for the helicity flux due to stretching of the
nonaxisymmetric magnetic field due to the differential rotation flux;
f)show the sum of the fluxes $F_{\alpha}$ and $F_{\beta}$ for the
nonaxisymmetric magnetic field.}
\end{figure}

Figure\ref{fig:tml} shows the time-latitude diagrams for the toroidal
magnetic field evolution, as well as , the small-scale helicity density,
$\left\langle \chi\right\rangle $ and the helicity density fluxes
$F_{\Omega}$, $F_{\alpha}$ and $F_{\beta}$ for axisymmetric and
nonaxisymmetric magnetic field. The model shows the regular dynamo
waves of the toroidal magnetic field which drifts toward equator in
course of the magnetic cycle. The emerging bipolar regions show no
effect on the butterfly diagram because the coupling between axisymmetric
and non-axisymmetric modes is weak. However, the bipolar regions have
the cumulative effect on the rate of the magnetic flux loss. Therefore
they affect the magnitude of the axisymmetric toroidal field. The
properties of the given model was discussed in details by \citet{Pipin2018d}.
The small-scale helicity density, $\left\langle \chi\right\rangle $
evolves in following conservation law helicity density. This conservation
law preserves the integral balance between $\left\langle \chi\right\rangle $,
$\overline{\mathbf{A}}\cdot\overline{\mathbf{B}}$ and $\overline{\tilde{\mathbf{A}}\cdot\tilde{\mathbf{B}}}$.
In the quasi stationary state the $\overline{\mathbf{A}}\cdot\overline{\mathbf{B}}$
contribution is much larger than $\overline{\tilde{\mathbf{A}}\cdot\tilde{\mathbf{B}}}$
(cf, Fig.\ref{fig:snap}d). Therefore, the $\left\langle \chi\right\rangle $
evolution follows the standard hemispheric helicity rule. The time-latitude
variations of the, $\overline{\tilde{\mathbf{A}}\cdot\tilde{\mathbf{B}}}=\overline{\tilde{\chi}}$,
show two bands in each hemisphere. The near equatorial bands show
the positive sign in the northern hemisphere and the negative in the
southern one. In the polar sides the situation is opposite. This patterns
results naturally from the longitudinal averaging of the synoptic
maps like that shown in Fig.\ref{fig:bip}c. Interesting that in the
run without tilt effect these bands show some equatorial drift, which
results into noisy behavior of the $\overline{\tilde{\mathbf{A}}\cdot\tilde{\mathbf{B}}}$
near equator in that run. The time-latitude diagram of the $\overline{\tilde{\mathbf{A}}\cdot\tilde{\mathbf{B}}}$
shows sometimes the same sign of helicity on both hemispheres. This
period correspond to the time when the wave of the toroidal magnetic
field goes close to equator. Therefore there is interaction of the
bipolar regions emerging in opposite hemispheres. The helicity fluxes
due to the differential rotation are opposite for the axisymmetric
and nonaxisymmetric magnetic field (see, Fig. \ref{fig:tml}b and
e). Our results show that the helicity fluxes due to the mean-electromotive
force effects are substantially less than those from effect of the
differential rotation. The butterfly diagram of the $F_{\Omega}$
agrees qualitatively with results of \citet{Hawkes2019AA} (see, Fig.3c,
there). 
\begin{figure}
\includegraphics[width=0.5\textwidth]{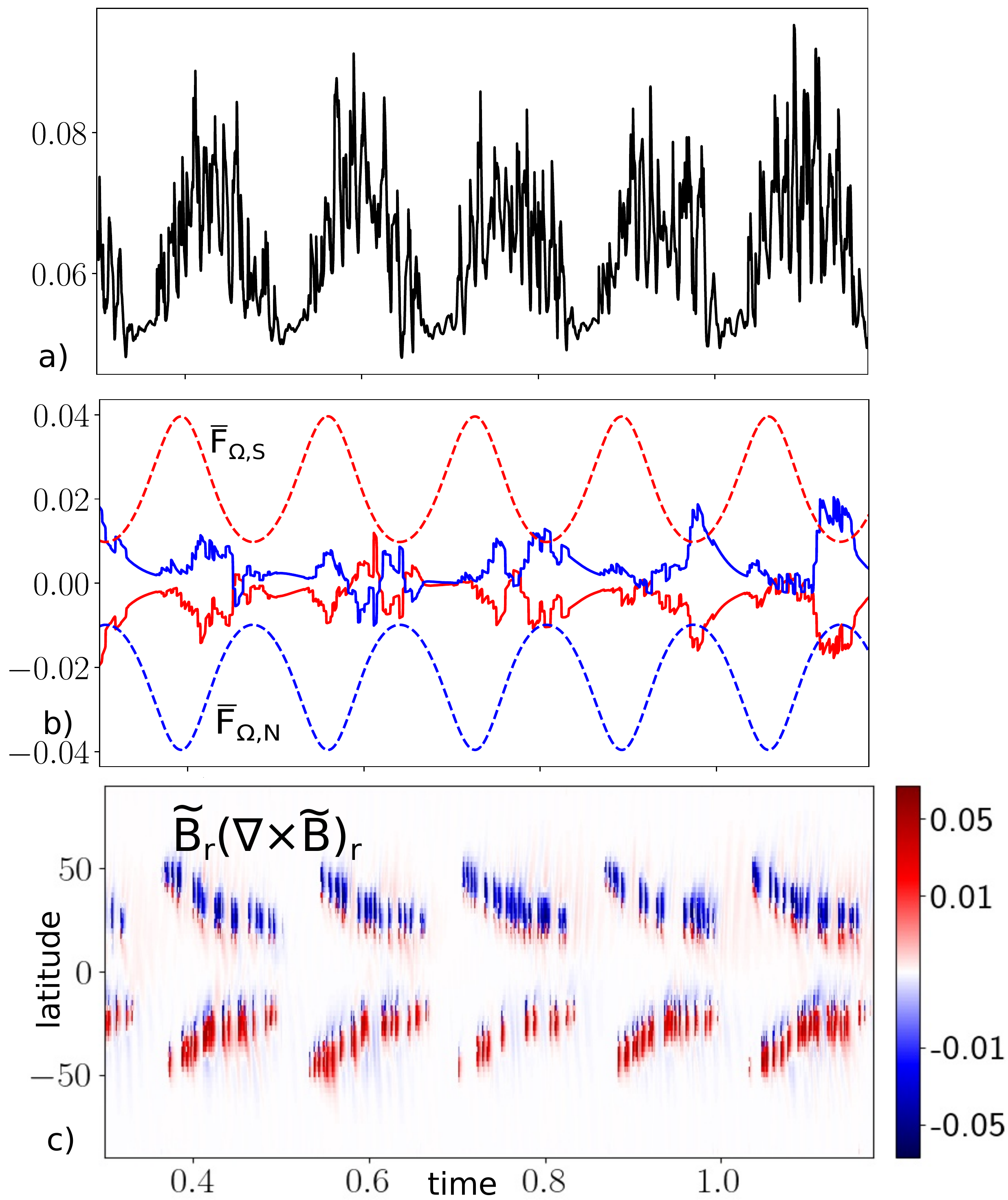}

\caption{\label{fig:int}a) The total flux of the radial magnetic field; b)the
axisymmetric (dashed lines) and nonaxisymmetric (solid lines) parts
of the helicity fluxes from differential rotation, for the North and
South hemispheres, c) the time-latitude evolution of the mean current
helicity generated by the emerging active regions.}

\end{figure}

Figure,\ref{fig:int} shows the integral parameters of the run  {and
variations of the current helicity density of the nonaxisymmetric
magnetic field.} The total flux of the radial magnetic field can be
considered as a proxy of the sunspot activity \citep{Stenflo2013a}.
Beside the main magnetic cycle variation, this parameter shows a short-term
quasi-biennial variations (see, \citealt{Frick2020}). In our model,
the quasi-biennial variations result from the evolution of the nonaxisymmetric
magnetic field, which is induced by the bipolar region formation.
The dynamo evolution of the axisymmetric and nonaxisymmetric magnetic
field results in the opposite helicity fluxes by the differential
rotation in the North and the South hemispheres. Note, the phase difference
between variations of the sunspot proxy, $\overline{F}_{\Omega}$
and $\tilde{F}_{\Omega}$. Our results agree qualitatively with the
solar observations (c.f., \citealt{Berger2000,Zhang2006a}). The  {current
helicity density of the nonaxisymmetric magnetic field show the standard
hemispheric helicity rule with the dominant negative sign in the North
hemisphere and the positive sign in the South hemisphere.}

\begin{figure}
\includegraphics[width=1\textwidth]{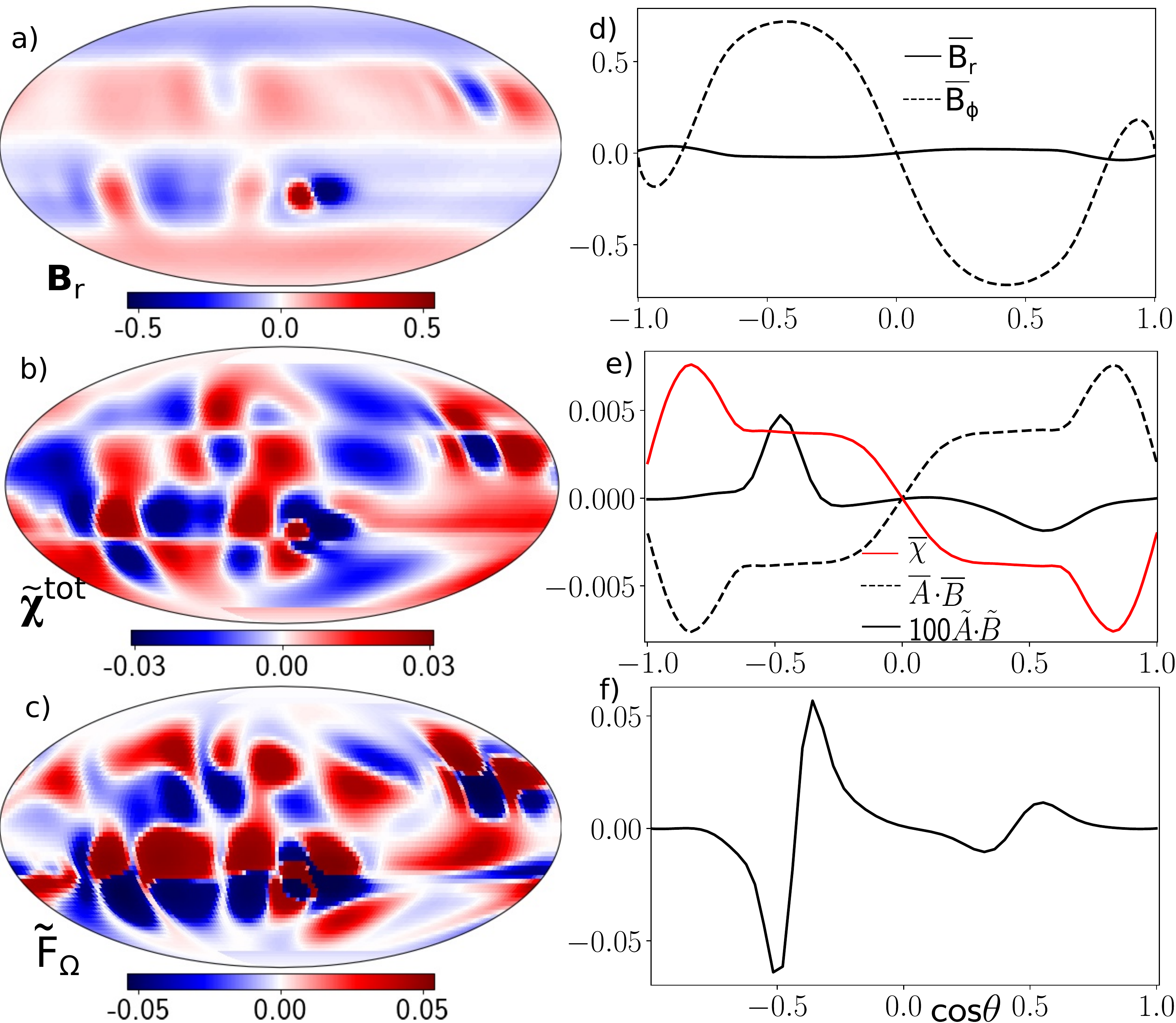}

\caption{\label{fig:snap}a) Snapshots of the radial magnetic field,  {$\left\langle B\right\rangle _{r}$,}
(color image) and the nonaxisymmetric toroidal field, $\tilde{B}_{\phi}$,
contours are shown in the same range as $\left\langle B\right\rangle _{r}$;
b) the total helicity density of the nonaxisymmetric field, $\tilde{\chi}^{(tot)}$,
(see, Eq.\ref{eq:t2}); c) the snapshot of the helicity density flux
due the differential rotation, $\tilde{F}_{\Omega}$(nonaxisymmetric
part); panels d),e) and f) show the mean latitudinal profiles of the
magnetic field and helicity parameters. The snapshots are taken around
the maximum of the toroidal magnetic field cycle, at the diffusive
time$\approx$1.}
\end{figure}

Figure \ref{fig:snap} shows snapshots of the magnetic field and the
helicity density distributions for the period of maximum of the toroidal
magnetic field cycle. { }The obtained magnetic field distribution
is much simpler than the longitudinal structure of the solar activity
proxies. This is likely because the range of the spatial scales which
are involved in the dynamo model is at least factor 10 less than in
the real Sun (see, Fig.10 in \citealt{Pipin2018d}). Also, the dynamo
model operates by the averaged equations, where, the effect of the
small-scale flows is replaced by the mean electromotive force. The
helicity patterns near the bipolar regions are qualitatively similar
to the case shown in Fig.\ref{fig:bip}b and not in Fig.\ref{fig:bip}c.
Note, that the Fig.\ref{fig:snap}b shows the helicity density of
the nonaxisymmetric magnetic field, $\tilde{\mathbf{A}}\cdot\tilde{\mathbf{B}}$.
The interaction of the emerging bi-poles with the background large-scale
nonaxisymmetric magnetic field produces the pattern similar to that
in Fig.\ref{fig:bip}b. At the right side of the snapshot we show
the mean distribution of the $\overline{\chi}$, $\overline{\mathbf{A}}\cdot\overline{\mathbf{B}}$
and $\overline{\tilde{\mathbf{A}}\cdot\tilde{\mathbf{B}}}$ for this
synoptic map. We find that in the model the $\overline{\tilde{\mathbf{A}}\cdot\tilde{\mathbf{B}}}$
has much less magnitude in compare with other two.

\section{Discussion and conclusions}

In the paper we study the effect of the emerging bipolar regions on
the magnetic helicity density distributions and the helicity density
fluxes. We employ a simplified dynamo model, which neglects the magnetic
evolution in the radial direction. In the past, the model of surface
dynamo waves was often used to illustrate the main principles and
basic effects of a dynamo operating in the thin shear layer \citep{Parker1993,Moss2004a,Moss2008}.
We generalize the toy model from 1D axisymmetric to 2D nonaxisymmetric
case in such a way that it keeps the basic properties of the axisymmetric
solar-type dynamo. Our formulation follows the framework of \citet{Moss2008}
and contains results of their model in 1D limit. These restrictions
result in the heuristic character of our model. Comparing our results
with observations, we have to bear in mind that in the model the magnetic
field dynamo generation and the bipolar region formation occur in
the same place. The formation of the bipolar regions by means of the
magnetic buoyancy instability is the main source of the nonaxisymmetric
magnetic field in our model. After \citet{Parker1984}, the magnetic
buoyancy instability is usually considered as a primary process in
sunspot formation. In the mean-field model, this instability results
in effective nonlinear pumping of the large-scale magnetic field outward
in radial direction. Our model employs this instability in the heuristic
way. Note, that the magnetic buoyancy instability is not the only
mechanism which is capable to form the sunspots from the large-scale
toroidal magnetic field (see, e.g., \citealt{Brandenburg2013}, and
a review of \citealt{Losada2017}). \citet{Pipin2018d} found that
the model shows the spectral properties of the nonaxisymmetric magnetic
field distribution in agreement with observations. They found that
the large-scale nonaxisymmetric magnetic field results mainly from
the diffusive decay of the bipolar regions. 

\begin{figure}
\includegraphics[width=0.7\textwidth]{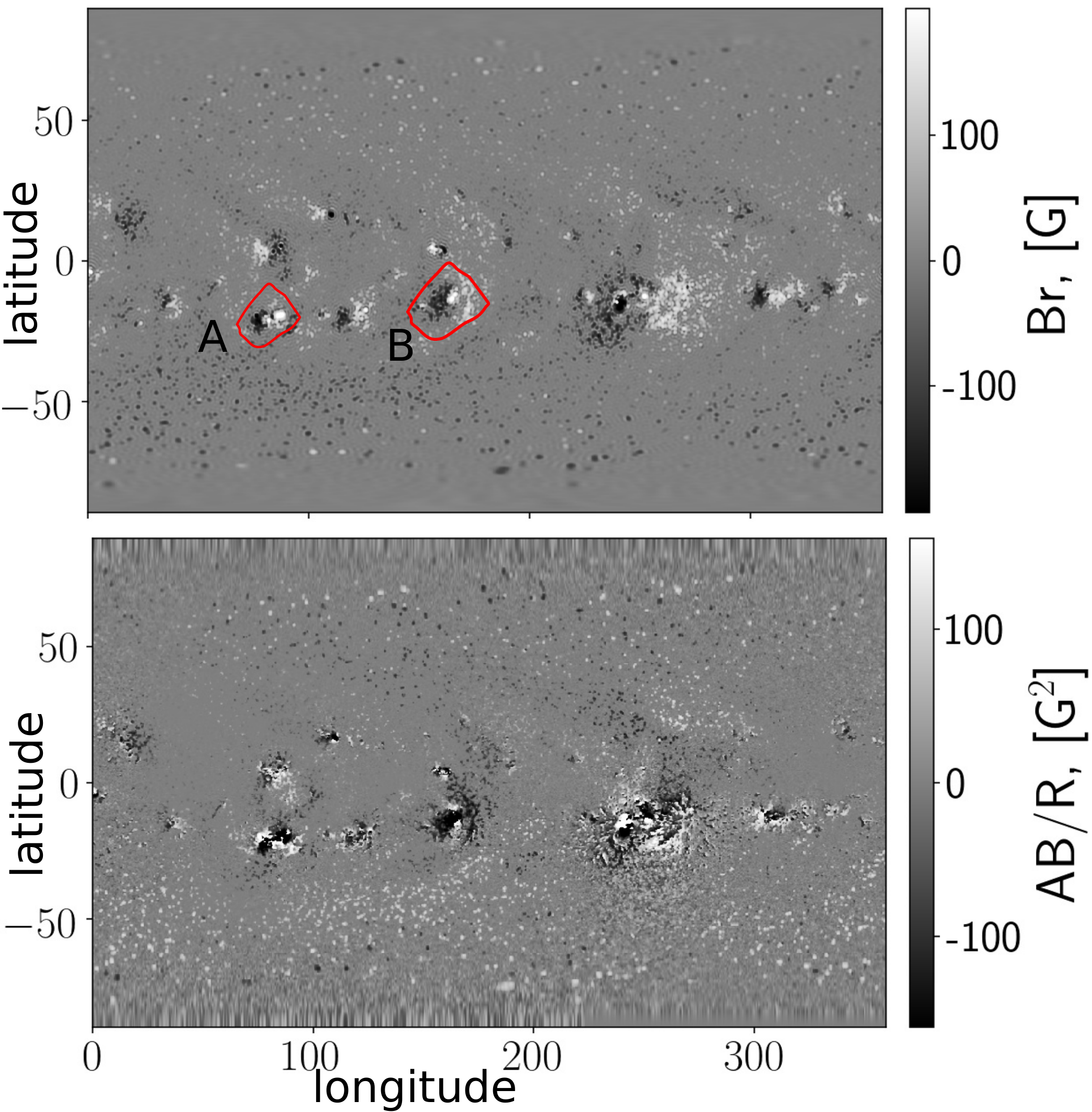}

\caption{\label{cr2157}Synoptic maps of the radial magnetic field (top) and
the magnetic helicity density (bottom) for the CR2157 (in following
to results of \citealt{Pipin2019a}). }
\end{figure}

In this paper our main goal was to find typical magnetic helicity
pattern which we can observe on the surface of the Sun. We look to
the case of the simple bipolar region which is formed from the large-scale
toroidal magnetic field by means of the magnetic buoyancy. We simulate
the tilt in heuristic way using the nonlinear $\alpha_{\beta}$-effect
which is induced by the magnetic buoyancy instability. The origin
of tilt is not specified. Such mechanisms was discussed in a number
of papers (see, \citealt{Leighton1969,Ferriz1994,Fisher1999,Charbonneau2011}).
We find that formation of the bipolar regions together with the large-scale
magnetic field produce the quadrupole magnetic helicity density pattern
(Figs\ref{fig:bip}b,\ref{fig:snap}c). The effect of the tilt in
the bipolar region results in the tilt of the helicity pattern. The
qualitatively similar results are found recently by \citet{Yeates2020ApJ}
using a completely different approach. \citet{Pipin2019a} studied
the helicity density distributions of the solar magnetic field using
observations of the Helioseismic and Magnetic Imager \citep[HMI, ][]{Scherrer2012}
on board Solar Dynamics Observatory \citep[SDO,][]{Pesnell2012}.
The Figure \ref{cr2157} shows examples of the magnetic field and
the magnetic helicity density synoptic maps for the Carrington rotation
2157. The two relatively small active regions in the southern hemisphere,
A and B show the quadrupole helicity density distributions in a qualitatively
agreement with results of our study. The best agreement is for the
region A. It is likely because it has a relatively simple distribution
of the radial magnetic field flux.

In our model the dynamo process and the bipolar region formation occur
in the same place. The emerging bipolar regions results in the loss
of the magnetic field flux and quenching the large-scale dynamo. We
find no a profound effect of these bi-poles on the helical properties
of the dynamo. This is likely due to specific of our model. Despite,
the nonlinear $\alpha_{\beta}$ effect produces some amount of the
large-scale poloidal magnetic field flux (see, e.g., Fig1a), it does
not affect much the background axisymmetric dynamo process. This is
contrary to assumptions of the Babcock-Leighton dynamo scenario \citep{Cameron2017}.
To satisfy this scenario we need a meridional circulation in the model.
This probably requires the 3D model. Besides, it is likely that the
large-scale dynamo, which utilizes such $\alpha_{\beta}$ effect acting
on the nonaxisymmetric magnetic field, operates in the highly supercritical
regimes \citep{Ferriz1994}. This is contrary to our original intention
to have the solar-type dynamo model as a limit in 1D-case. Therefore,
we can not expect a considerable effect of the generated $\overline{\tilde{\mathbf{A}}\cdot\tilde{\mathbf{B}}}$
on the small-scale helicity density evolution. The modeled $\left\langle \chi\right\rangle $
distributions are determined by the axisymmetric type of the dynamo
model. The $\overline{\tilde{\mathbf{A}}\cdot\tilde{\mathbf{B}}}$
shows the inverted HHR near equator in compare the sign of the current
helicity density of the solar active regions, \citep{Zhang2010,Zhang2016}.
The positive sign of the magnetic helicity density from tilted bipolar
regions was anticipated in earlier studies \citep{Pevtsov2014}. Theoretically,
it is expected that the tilt can results in the internal twist opposing
to the writhe by the magnetic tensions \citep{Blackman2003}. In our
model this process is taken into account by the conservation law.
Following to this law the small-scale helicity density $\overline{\chi}$
evolves in balance with the $\overline{\tilde{\mathbf{A}}\cdot\tilde{\mathbf{B}}}$
helicity density of the axisymmetric field, $\overline{\mathbf{A}}\cdot\overline{\mathbf{B}}$.
The heuristic type of our model do not allow a detailed quantitative
comparison with observations. This can be improved in the 3D models,
which allow to take into account all the dynamo effects in a physically
consistent way. We may anticipate that for a shallow process of the
sunspot formation, the 3D models can give the results which agrees
qualitatively with ours. 

Summing up, it is found the emerging bipolar regions produce the quadrupole
helicity density patterns. The similar patterns were found for the
helicity flux by means of the differential rotation. We find that
the tilted bipolar regions show the inverted hemispheric helicity
rule near equator in compare with observations of the magnetic helicity
in the solar active regions. In general, our results suggest that
on the intermediate scales such as the scale of the bipolar active
regions the averaged magnetic helicity distribution can show no definite
sign distribution in the northern and southern hemisphere of the Sun.

 {Acknowledgements} The author acknowledge the financial support
by the Russian Foundation for Basic Research grant 19-52-53045 and
support of scientific project FR II.16 of ISTP SB RAS. This work utilizes
HMI data which are used here are courtesy of NASA/SDO and the HMI
science teams. Some part of this work was presented and discussed
during ``Solar Helicities in Theory and Observations: Implications
for Space Weather and Dynamo Theory'' Program at Nordic Institute
for Theoretical Physics (NORDITA) in 4--29 March 2019. I thank the
anonymous referee for the helpful comments, and constructive remarks
on this manuscript. 

\bibliographystyle{jpp}

\begin{thebibliography}{57}
\expandafter\ifx\csname natexlab\endcsname\relax\def\natexlab#1{#1}\fi
\def\au#1{#1} \def\ed#1{#1} \def\yr#1{#1}\def\at#1{#1}\def\jt#1{\textit{#1}}
  \def\bt#1{#1}\def\bvol#1{\textbf{#1}} \def\vol#1{#1} \def\pg#1{#1}
  \def\publ#1{#1}\def\arxiv#1{#1}\def\org#1{#1}\def\st#1{\textit{#1}}

\bibitem[{Bao} \& {Zhang}(1998)]{Bao1998}
{\sc \au{{Bao}, S.} \& \au{{Zhang}, H.}} \yr{1998}  \at{Patterns of current
  helicity for the twenty-second solar}.  \jt{\apj}  \bvol{496},
  \pg{L43--L46}.

\bibitem[Berger(1984)]{Berger1984}
{\sc \au{Berger, Mitchell A.;~Field, G.~B.}} \yr{1984}  \at{The topological
  properties of magnetic helicity}.  \jt{Journal of Fluid Mechanics}
  \bvol{147}.

\bibitem[{Berger} \& {Hornig}(2018)]{Berger2018}
{\sc \au{{Berger}, M.~A.} \& \au{{Hornig}, G.}} \yr{2018}  \at{A generalized
  poloidal-toroidal decomposition and an absolute measure of helicity}.
  \jt{Journal of Physics A Mathematical General}  \bvol{51},  \pg{495501}.

\bibitem[{Berger} \& {Ruzmaikin}(2000)]{Berger2000}
{\sc \au{{Berger}, M.~A.} \& \au{{Ruzmaikin}, A.}} \yr{2000}  \at{Rate of
  helicity production by solar rotation}.  \jt{\jgr}  \bvol{105},
  \pg{10481--10490}.

\bibitem[{Bigazzi} \& {Ruzmaikin}(2004)]{Bigazzi2004}
{\sc \au{{Bigazzi}, A.} \& \au{{Ruzmaikin}, A.}} \yr{2004}  \at{The sun's
  preferred longitudes and the coupling of magnetic dynamo modes}.  \jt{\apj}
  \bvol{604},  \pg{944--959},  \arxiv{arXiv: astro-ph/0312212}.

\bibitem[{Blackman} \& {Brandenburg}(2003)]{Blackman2003}
{\sc \au{{Blackman}, E.~G.} \& \au{{Brandenburg}, A.}} \yr{2003}  \at{Doubly
  helical coronal ejections from dynamos and their role in sustaining the solar
  cycle}.  \jt{\apjl}  \bvol{584},  \pg{L99--L102},  \arxiv{arXiv:
  astro-ph/0212010}.

\bibitem[{Blackman} \& {Thomas}(2015)]{Blackman2015}
{\sc \au{{Blackman}, E.~G.} \& \au{{Thomas}, J.~H.}} \yr{2015}  \at{Explaining
  the observed relation between stellar activity and rotation}.  \jt{\mnras}
  \bvol{446},  \pg{L51--L55},  \arxiv{arXiv: 1407.8500}.

\bibitem[{Brandenburg}(2018)]{Brandenburg2018}
{\sc \au{{Brandenburg}, A.}} \yr{2018}  \at{Advances in mean-field dynamo
  theory and applications to astrophysical turbulence}.  \jt{Journal of Plasma
  Physics}  \bvol{84},  \pg{735840404}.

\bibitem[{Brandenburg} {\em et~al.\/}(2013){Brandenburg}, {Kleeorin} \&
  {Rogachevskii}]{Brandenburg2013}
{\sc \au{{Brandenburg}, A.}, \au{{Kleeorin}, N.} \& \au{{Rogachevskii}, I.}}
  \yr{2013}  \at{Self-assembly of shallow magnetic spots through strongly
  stratified turbulence}.  \jt{\apj}  \bvol{776},  \pg{L23},  \arxiv{arXiv:
  1306.4915}.

\bibitem[{Brandenburg} {\em et~al.\/}(2017){Brandenburg}, {Petrie} \&
  {Singh}]{Brandenburg2017}
{\sc \au{{Brandenburg}, A.}, \au{{Petrie}, G. J.~D.} \& \au{{Singh}, N.~K.}}
  \yr{2017}  \at{Two-scale analysis of solar magnetic helicity}.  \jt{\apj}
  \bvol{836},  \pg{21}.

\bibitem[{Brandenburg} \& {Subramanian}(2005)]{Brandenburg2005b}
{\sc \au{{Brandenburg}, A.} \& \au{{Subramanian}, K.}} \yr{2005}
  \at{Astrophysical magnetic fields and nonlinear dynamo theory}.
  \jt{\physrep}  \bvol{417},  \pg{1--209},  \arxiv{arXiv:
  arXiv:astro-ph/0405052}.

\bibitem[{Cameron} \& {Sch{\"u}ssler}(2017{\natexlab{{\em a\/}}})]{Cameron17}
{\sc \au{{Cameron}, R.~H.} \& \au{{Sch{\"u}ssler}, M.}} \yr{2017{\natexlab{{\em
  a\/}}}}  \at{{An update of Leighton's solar dynamo model}}.  \jt{\aap}
  \bvol{599},  \pg{A52},  \arxiv{arXiv: 1611.09111}.

\bibitem[{Cameron} \& {Sch{\"u}ssler}(2017{\natexlab{{\em b\/}}})]{Cameron2017}
{\sc \au{{Cameron}, R.~H.} \& \au{{Sch{\"u}ssler}, M.}} \yr{2017{\natexlab{{\em
  b\/}}}}  \at{Understanding solar cycle variability}.  \jt{\apj}  \bvol{843},
  \pg{111},  \arxiv{arXiv: 1705.10746}.

\bibitem[{Charbonneau}(2011)]{Charbonneau2011}
{\sc \au{{Charbonneau}, P.}} \yr{2011}  \at{Dynamo models of the solar cycle}.
  \jt{Living Reviews in Solar Physics}  \bvol{2},  \pg{2}.

\bibitem[{Ferriz-Mas} {\em et~al.\/}(1994){Ferriz-Mas}, {Schmitt} \&
  {Schuessler}]{Ferriz1994}
{\sc \au{{Ferriz-Mas}, A.}, \au{{Schmitt}, D.} \& \au{{Schuessler}, M.}}
  \yr{1994}  \at{{A dynamo effect due to instability of magnetic flux tubes.}}
  \jt{\aap}  \bvol{289},  \pg{949--956}.

\bibitem[{Fisher} {\em et~al.\/}(1999){Fisher}, {Longcope}, {Linton}, {Fan} \&
  {Pevtsov}]{Fisher1999}
{\sc \au{{Fisher}, G.~H.}, \au{{Longcope}, D.~W.}, \au{{Linton}, M.~G.},
  \au{{Fan}, Y.} \& \au{{Pevtsov}, A.~A.}} \yr{1999} The origin and role of
  twist in active regions.  \bt{In {\em Stellar Dynamos: Nonlinearity and
  Chaotic Flows\/} (ed. \ed{M.~{Nunez} \& A.~{Ferriz-Mas}})},  \st{Astronomical
  Society of the Pacific Conference Series},  \vol{vol. 178},  \pg{p.~35}.

\bibitem[{Frick} {\em et~al.\/}(2020){Frick}, {Sokoloff}, {Stepanov}, {Pipin}
  \& {Usoskin}]{Frick2020}
{\sc \au{{Frick}, P.}, \au{{Sokoloff}, D.}, \au{{Stepanov}, R.}, \au{{Pipin},
  V.} \& \au{{Usoskin}, I.}} \yr{2020}  \at{{Spectral characteristic of
  mid-term quasi-periodicities in sunspot data}}.  \jt{\mnras}  \bvol{491}~(4),
   \pg{5572--5578},  \arxiv{arXiv: 1911.06881}.

\bibitem[{Hawkes} \& {Yeates}(2019)]{Hawkes2019AA}
{\sc \au{{Hawkes}, G.} \& \au{{Yeates}, A.~R.}} \yr{2019}  \at{{Hemispheric
  injection of magnetic helicity by surface flux transport}}.  \jt{\aap}
  \bvol{631},  \pg{A138}.

\bibitem[{Hubbard} \& {Brandenburg}(2012)]{Hubbard2012}
{\sc \au{{Hubbard}, A.} \& \au{{Brandenburg}, A.}} \yr{2012}  \at{Catastrophic
  quenching in {$\alpha$}{$\Omega$} dynamos revisited}.  \jt{\apj}  \bvol{748},
   \pg{51},  \arxiv{arXiv: 1107.0238}.

\bibitem[{Jennings} {\em et~al.\/}(1990){Jennings}, {Brandenburg}, {Tuominen}
  \& {Moss}]{Jennings1990}
{\sc \au{{Jennings}, R.}, \au{{Brandenburg}, A.}, \au{{Tuominen}, I.} \&
  \au{{Moss}, D.}} \yr{1990}  \at{Can stellar dynamos be modelled in less than
  three dimensions?}  \jt{\aap}  \bvol{230},  \pg{463--473}.

\bibitem[{Kitchatinov} \& {Pipin}(1993)]{Kitchatinov1993}
{\sc \au{{Kitchatinov}, L.~L.} \& \au{{Pipin}, V.~V.}} \yr{1993}
  \at{Mean-field buoyancy}.  \jt{\aap}  \bvol{274},  \pg{647--652}.

\bibitem[{Kleeorin} {\em et~al.\/}(2000){Kleeorin}, {Moss}, {Rogachevskii} \&
  {Sokoloff}]{Kleeorin2000}
{\sc \au{{Kleeorin}, N.}, \au{{Moss}, D.}, \au{{Rogachevskii}, I.} \&
  \au{{Sokoloff}, D.}} \yr{2000}  \at{Helicity balance and steady-state
  strength of the dynamo generated galactic magnetic field}.  \jt{\aap}
  \bvol{361},  \pg{L5--L8},  \arxiv{arXiv: arXiv:astro-ph/0205266}.

\bibitem[Kleeorin \& Rogachevskii(1999)]{Kleeorin1999}
{\sc \au{Kleeorin, N.} \& \au{Rogachevskii, I.}} \yr{1999}  \at{Magnetic
  helicity tensor for an anisotropic turbulence}.  \jt{Phys. Rev.E}  \bvol{59},
   \pg{6724--6729}.

\bibitem[Kleeorin \& Ruzmaikin(1982)]{Kleeorin1982}
{\sc \au{Kleeorin, N.~I.} \& \au{Ruzmaikin, A.~A.}} \yr{1982}  \at{Dynamics of
  the average turbulent helicity in a magnetic field}.
  \jt{Magnetohydrodynamics}  \bvol{18},  \pg{116--122}.

\bibitem[Krause \& R\"adler(1980)]{Krause1980}
{\sc \au{Krause, F.} \& \au{R\"adler, K.-H.}} \yr{1980} {\em Mean-Field
  Magnetohydrodynamics and Dynamo Theory\/}.  \publ{Berlin: Akademie-Verlag}.

\bibitem[{Leighton}(1969)]{Leighton1969}
{\sc \au{{Leighton}, R.~B.}} \yr{1969}  \at{{A Magneto-Kinematic Model of the
  Solar Cycle}}.  \jt{\apj}  \bvol{156},  \pg{1}.

\bibitem[{Losada} {\em et~al.\/}(2017){Losada}, {Warnecke}, {Glogowski},
  {Roth}, {Brandenburg}, {Kleeorin} \& {Rogachevskii}]{Losada2017}
{\sc \au{{Losada}, I.~R.}, \au{{Warnecke}, J.}, \au{{Glogowski}, K.},
  \au{{Roth}, M.}, \au{{Brandenburg}, A.}, \au{{Kleeorin}, N.} \&
  \au{{Rogachevskii}, I.}} \yr{2017} A new look at sunspot formation using
  theory and observations.  \bt{In {\em Fine Structure and Dynamics of the
  Solar Atmosphere\/} (ed. \ed{S.~{Vargas Dom{\'{\i}}nguez}, A.~G.
  {Kosovichev}, P.~{Antolin} \& L.~{Harra}})},  \st{IAU Symposium},  \vol{vol.
  327},  \pg{pp. 46--59},  \arxiv{arXiv: 1704.04062}.

\bibitem[{Mackay} \& {Yeates}(2012)]{Mackay2012}
{\sc \au{{Mackay}, D.~H.} \& \au{{Yeates}, A.~R.}} \yr{2012}  \at{The sun's
  global photospheric and coronal magnetic fields: Observations and models}.
  \jt{Living Reviews in Solar Physics}  \bvol{9},  \pg{6},  \arxiv{arXiv:
  1211.6545}.

\bibitem[{Mitra} {\em et~al.\/}(2010){Mitra}, {Candelaresi}, {Chatterjee},
  {Tavakol} \& {Brandenburg}]{Mitra2010}
{\sc \au{{Mitra}, D.}, \au{{Candelaresi}, S.}, \au{{Chatterjee}, P.},
  \au{{Tavakol}, R.} \& \au{{Brandenburg}, A.}} \yr{2010}  \at{Equatorial
  magnetic helicity flux in simulations with different gauges}.
  \jt{Astronomische Nachrichten}  \bvol{331},  \pg{130},  \arxiv{arXiv:
  0911.0969}.

\bibitem[Moffatt(1978)]{Moffatt1978}
{\sc \au{Moffatt, H.~K.}} \yr{1978} {\em Magnetic Field Generation in
  Electrically Conducting Fluids\/}.  \publ{Cambridge, England: Cambridge
  University Press}.

\bibitem[{Moss} {\em et~al.\/}(2004){Moss}, {Sokoloff}, {Kuzanyan} \&
  {Petrov}]{Moss2004a}
{\sc \au{{Moss}, D.}, \au{{Sokoloff}, D.}, \au{{Kuzanyan}, K.} \& \au{{Petrov},
  A.}} \yr{2004}  \at{Stellar dynamo waves: Asymptotic configurations}.
  \jt{Geophysical and Astrophysical Fluid Dynamics}  \bvol{98},  \pg{257--272}.

\bibitem[{Moss} {\em et~al.\/}(2008){Moss}, {Sokoloff}, {Usoskin} \&
  {Tutubalin}]{Moss2008}
{\sc \au{{Moss}, D.}, \au{{Sokoloff}, D.}, \au{{Usoskin}, I.} \&
  \au{{Tutubalin}, V.}} \yr{2008}  \at{Solar grand minima and random
  fluctuations in dynamo parameters}.  \jt{Solar Phys.}  \bvol{250},
  \pg{221--234}.

\bibitem[{Noyes} {\em et~al.\/}(1984){Noyes}, {Weiss} \& {Vaughan}]{Noyes1984}
{\sc \au{{Noyes}, R.~W.}, \au{{Weiss}, N.~O.} \& \au{{Vaughan}, A.~H.}}
  \yr{1984}  \at{The relation between stellar rotation rate and activity cycle
  periods}.  \jt{\apj}  \bvol{287},  \pg{769--773}.

\bibitem[{Parker}(1984)]{Parker1984}
{\sc \au{{Parker}, E.~N.}} \yr{1984}  \at{Magnetic buoyancy and the escape of
  magnetic fields from stars}.  \jt{\apj}  \bvol{281},  \pg{839--845}.

\bibitem[{Parker}(1993)]{Parker1993}
{\sc \au{{Parker}, E.~N.}} \yr{1993}  \at{A solar dynamo surface wave at the
  interface between convection and nonuniform rotation}.  \jt{\apj}
  \bvol{408},  \pg{707--719}.

\bibitem[{Pesnell} {\em et~al.\/}(2012){Pesnell}, {Thompson} \&
  {Chamberlin}]{Pesnell2012}
{\sc \au{{Pesnell}, W.~D.}, \au{{Thompson}, B.~J.} \& \au{{Chamberlin}, P.~C.}}
  \yr{2012}  \at{The solar dynamics observatory (sdo)}.  \jt{\solphys}
  \bvol{275},  \pg{3--15}.

\bibitem[{Pevtsov} {\em et~al.\/}(2014){Pevtsov}, {Berger}, {Nindos}, {Norton}
  \& {van Driel-Gesztelyi}]{Pevtsov2014}
{\sc \au{{Pevtsov}, A.~A.}, \au{{Berger}, M.~A.}, \au{{Nindos}, A.},
  \au{{Norton}, A.~A.} \& \au{{van Driel-Gesztelyi}, L.}} \yr{2014}
  \at{Magnetic helicity, tilt, and twist}.  \jt{\ssr}  \bvol{186},
  \pg{285--324}.

\bibitem[{Pevtsov} {\em et~al.\/}(1994){Pevtsov}, {Canfield} \&
  {Metcalf}]{Pevtsov1994}
{\sc \au{{Pevtsov}, A.~A.}, \au{{Canfield}, R.~C.} \& \au{{Metcalf}, T.~R.}}
  \yr{1994}  \at{Patterns of helicity in solar active regions}.  \jt{\apjl}
  \bvol{425},  \pg{L117--L119}.

\bibitem[Pipin(2018)]{Pipin2018}
{\sc \au{Pipin, V.}} \yr{2018} Vvpipin/2dspdy 0.1.1.

\bibitem[Pipin \& Kosovichev(2015)]{Pipin2015a}
{\sc \au{Pipin, V.~V.} \& \au{Kosovichev, A.~G.}} \yr{2015}  \at{Effects of
  large-scale non-axisymmetric perturbations in the mean-field solar dynamo.}
  \jt{The Astrophysical Journal}  \bvol{813}~(2),  \pg{134}.

\bibitem[{Pipin} \& {Kosovichev}(2018)]{Pipin2018d}
{\sc \au{{Pipin}, V.~V.} \& \au{{Kosovichev}, A.~G.}} \yr{2018}  \at{Does
  nonaxisymmetric dynamo operate in the sun?}  \jt{ArXiv e-prints}  \pg{p.
  arXiv:1808.05332},  \arxiv{arXiv: 1808.05332}.

\bibitem[{Pipin} \& {Pevtsov}(2014)]{Pipin2014b}
{\sc \au{{Pipin}, V.~V.} \& \au{{Pevtsov}, A.~A.}} \yr{2014}  \at{Magnetic
  helicity of the global field in solar cycles 23 and 24}.  \jt{\apj}
  \bvol{789},  \pg{21},  \arxiv{arXiv: 1402.2386}.

\bibitem[{Pipin} {\em et~al.\/}(2019){Pipin}, {Pevtsov}, {Liu} \&
  {Kosovichev}]{Pipin2019a}
{\sc \au{{Pipin}, V.~V.}, \au{{Pevtsov}, A.~A.}, \au{{Liu}, Y.} \&
  \au{{Kosovichev}, A.~G.}} \yr{2019}  \at{Evolution of magnetic helicity in
  solar cycle 24}.  \jt{\apj}  \bvol{877}~(2),  \pg{L36},  \arxiv{arXiv:
  1905.00772}.

\bibitem[{Pipin} {\em et~al.\/}(2013){Pipin}, {Sokoloff}, {Zhang} \&
  {Kuzanyan}]{Pipin2013c}
{\sc \au{{Pipin}, V.~V.}, \au{{Sokoloff}, D.~D.}, \au{{Zhang}, H.} \&
  \au{{Kuzanyan}, K.~M.}} \yr{2013}  \at{Helicity conservation in nonlinear
  mean-field solar dynamo}.  \jt{\apj}  \bvol{768},  \pg{46},  \arxiv{arXiv:
  1211.2420}.

\bibitem[Pouquet {\em et~al.\/}(1975)Pouquet, Frisch \& L\'eorat]{Pouquet1975}
{\sc \au{Pouquet, A.}, \au{Frisch, U.} \& \au{L\'eorat, J.}} \yr{1975}
  \at{Strong {MHD} helical turbulence and the nonlinear dynamo effect}.  \jt{J.
  Fluid Mech.}  \bvol{68},  \pg{769--778}.

\bibitem[{R{\"u}diger} {\em et~al.\/}(2011){R{\"u}diger}, {Kitchatinov} \&
  {Brandenburg}]{Ruediger2011}
{\sc \au{{R{\"u}diger}, G.}, \au{{Kitchatinov}, L.~L.} \& \au{{Brandenburg},
  A.}} \yr{2011}  \at{Cross helicity and turbulent magnetic diffusivity in the
  solar convection zone}.  \jt{\solphys}  \bvol{269},  \pg{3--12},
  \arxiv{arXiv: 1004.4881}.

\bibitem[{Ruediger} \& {Kichatinov}(1993)]{Ruediger1993b}
{\sc \au{{Ruediger}, G.} \& \au{{Kichatinov}, L.~L.}} \yr{1993}
  \at{{Alpha-effect and alpha-quenching}}.  \jt{\aap}  \bvol{269}~(1-2),
  \pg{581--588}.

\bibitem[{Scherrer} {\em et~al.\/}(2012){Scherrer}, {Schou}, {Bush},
  {Kosovichev}, {Bogart}, {Hoeksema}, {Liu}, {Duvall}, {Zhao}, {Title},
  {Schrijver}, {Tarbell} \& {Tomczyk}]{Scherrer2012}
{\sc \au{{Scherrer}, P.~H.}, \au{{Schou}, J.}, \au{{Bush}, R.~I.},
  \au{{Kosovichev}, A.~G.}, \au{{Bogart}, R.~S.}, \au{{Hoeksema}, J.~T.},
  \au{{Liu}, Y.}, \au{{Duvall}, T.~L.}, \au{{Zhao}, J.}, \au{{Title}, A.~M.},
  \au{{Schrijver}, C.~J.}, \au{{Tarbell}, T.~D.} \& \au{{Tomczyk}, S.}}
  \yr{2012}  \at{The helioseismic and magnetic imager (hmi) investigation for
  the solar dynamics observatory (sdo)}.  \jt{\solphys}  \bvol{275},
  \pg{207--227}.

\bibitem[{Seehafer}(1994)]{Seehafer1994}
{\sc \au{{Seehafer}, N.}} \yr{1994}  \at{Alpha effect in the solar atmosphere}.
   \jt{\aap}  \bvol{284},  \pg{593--598}.

\bibitem[{Singh} {\em et~al.\/}(2018){Singh}, {K{\"a}pyl{\"a}}, {Brandenburg},
  {K{\"a}pyl{\"a}}, {Lagg} \& {Virtanen}]{Singh2018}
{\sc \au{{Singh}, N.~K.}, \au{{K{\"a}pyl{\"a}}, M.~J.}, \au{{Brandenburg}, A.},
  \au{{K{\"a}pyl{\"a}}, Petri, J.}, \au{{Lagg}, A.} \& \au{{Virtanen}, I.}}
  \yr{2018}  \at{Bihelical spectrum of solar magnetic helicity and its
  evolution}.  \jt{\apj}  \bvol{863},  \pg{182},  \arxiv{arXiv: 1804.04994}.

\bibitem[{Stenflo}(2013)]{Stenflo2013a}
{\sc \au{{Stenflo}, J.~O.}} \yr{2013}  \at{Solar magnetic fields as revealed by
  stokes polarimetry}.  \jt{\aapr}  \bvol{21},  \pg{66},  \arxiv{arXiv:
  1309.5454}.

\bibitem[{Tlatov} {\em et~al.\/}(2013){Tlatov}, {Illarionov}, {Sokoloff} \&
  {Pipin}]{Tlatov2013}
{\sc \au{{Tlatov}, A.}, \au{{Illarionov}, E.}, \au{{Sokoloff}, D.} \&
  \au{{Pipin}, V.}} \yr{2013}  \at{{A new dynamo pattern revealed by the tilt
  angle of bipolar sunspot groups}}.  \jt{\mnras}  \bvol{432}~(4),
  \pg{2975--2984},  \arxiv{arXiv: 1302.2715}.

\bibitem[{Toriumi} \& {Wang}(2019)]{Toriumi2019}
{\sc \au{{Toriumi}, S.} \& \au{{Wang}, H.}} \yr{2019}  \at{Flare-productive
  active regions}.  \jt{Living Reviews in Solar Physics}  \bvol{16}~(1),
  \pg{3},  \arxiv{arXiv: 1904.12027}.

\bibitem[{Yeates}(2020)]{Yeates2020ApJ}
{\sc \au{{Yeates}, A.~R.}} \yr{2020}  \at{{The Minimal Helicity of Solar
  Coronal Magnetic Fields}}.  \jt{\apjl}  \bvol{898}~(2),  \pg{L49},
  \arxiv{arXiv: 2007.10649}.

\bibitem[{Zhang} {\em et~al.\/}(2016){Zhang}, {Brandenburg} \&
  {Sokoloff}]{Zhang2016}
{\sc \au{{Zhang}, H.}, \au{{Brandenburg}, A.} \& \au{{Sokoloff}, D.~D.}}
  \yr{2016}  \at{Evolution of magnetic helicity and energy spectra of solar
  active regions}.  \jt{\apj}  \bvol{819},  \pg{146}.

\bibitem[{Zhang} {\em et~al.\/}(2010){Zhang}, {Sakurai}, {Pevtsov}, {Gao},
  {Xu}, {Sokoloff} \& {Kuzanyan}]{Zhang2010}
{\sc \au{{Zhang}, H.}, \au{{Sakurai}, T.}, \au{{Pevtsov}, A.}, \au{{Gao}, Y.},
  \au{{Xu}, H.}, \au{{Sokoloff}, D.~D.} \& \au{{Kuzanyan}, K.}} \yr{2010}
  \at{A new dynamo pattern revealed by solar helical magnetic fields}.
  \jt{\mnras}  \bvol{402},  \pg{L30--L33},  \arxiv{arXiv: 0911.5713}.

\bibitem[{Zhang}(2006)]{Zhang2006a}
{\sc \au{{Zhang}, M.}} \yr{2006}  \at{Helicity observations of weak and strong
  fields}.  \jt{\apjl}  \bvol{646},  \pg{L85--L88},  \arxiv{arXiv:
  astro-ph/0606231}.

\end{thebibliography}

\section{Appendix}

\subsection{The large-scale magnetic field and its vector potential\label{supp}}

We decompose the total magnetic field induction vector on the sum
of the axisymmetric and non-axisymmetric parts:$\left\langle \mathbf{B}\right\rangle =\overline{\mathbf{B}}+\tilde{\mathbf{B}}$.
In the spherical coordinates, the axisymmetric part, $\overline{\mathbf{B}}$,
is represented as follows:
\begin{eqnarray}
\mathbf{\overline{B}} & = & \hat{\mathbf{\phi}}B+\nabla\times\left(A\hat{\mathbf{\phi}}\right)\\
 & = & \hat{\mathbf{\phi}}B-\frac{\hat{r}}{r}\frac{\partial A\sin\theta}{\partial\mu}-\frac{\hat{\theta}}{r}\frac{\partial rA}{\partial r},\nonumber 
\end{eqnarray}
where, $\hat{r}$ is the unit vector in the radial direction, $\hat{\theta}$
is the unit vector in meridional direction, and $\hat{\phi}$ is the
unit vector in the azimuthal direction, $\mu=\cos\theta$. In this
paper we assume that the scalars $A$ and $B$ are independent of
radius. Therefore, for the axisymmetric magnetic field at $r=R$ we
get

\begin{eqnarray}
\mathbf{\overline{B}} & = & \hat{\mathbf{\phi}}B-\frac{\hat{r}}{R}\frac{\partial A\sin\theta}{\partial\mu}-\frac{\hat{\theta}A}{R}.
\end{eqnarray}
The above definitions preserve the divergency free vector-field, $\boldsymbol{\nabla}\cdot\mathbf{\overline{B}}=0$.
For the axisymmetric part of the vector potential we have
\begin{equation}
\overline{\mathbf{A}}=\hat{r}A_{r}+\hat{\mathbf{\phi}}A,\label{eq:app}
\end{equation}
where $B=\partial A_{r}/\partial\mu$ (see, \citealt{Pipin2014b}).
The unique axisymmetric potential is found for the gauge $\int_{-1}^{1}A_{r}d\mu=0$
(cf., below derivations). For the non-axisymmetric part of magnetic
field, we use the same idea. In following \citet{Krause1980}, we
write:

\begin{eqnarray}
\mathbf{\tilde{B}} & = & \boldsymbol{\nabla}\times\left(\hat{r}T\right)+\boldsymbol{\nabla}\times\boldsymbol{\nabla}\times\left(\hat{r}S\right)\label{eq:bdec}\\
 & = & -\frac{\hat{r}}{r}\Delta_{\Omega}S+\hat{\theta}\left(\frac{1}{\sin\theta}\frac{\partial T}{\partial\phi}-\frac{\sin\theta}{r}\frac{\partial}{\partial\mu}\frac{\partial rS}{\partial r}\right)+\hat{\phi}\left(\sin\theta\frac{\partial T}{\partial\mu}+\frac{1}{r\sin\theta}\frac{\partial}{\partial\phi}\frac{\partial rS}{\partial r}\right),\nonumber 
\end{eqnarray}
where ${\displaystyle \Delta_{\Omega}=\frac{\partial}{\partial\mu}\sin^{2}\theta\frac{\partial}{\partial\mu}+\frac{1}{\sin^{2}\theta}\frac{\partial^{2}}{\partial\phi^{2}}}$.
Besides, we apply the following gauge (see, e.g., \citealp{Krause1980}):
\begin{equation}
\int_{0}^{2\pi}\int_{-1}^{1}Sd\mu d\phi=\int_{0}^{2\pi}\int_{-1}^{1}Td\mu d\phi=0.\label{eq:norm}
\end{equation}
Assuming that the potential's scalar functions $S$ and $T$ are independent
of the radius, we get, 

\begin{equation}
\mathbf{\tilde{B}}=-\frac{\hat{r}}{R}\Delta_{\Omega}S+\hat{\theta}\left(\frac{1}{\sin\theta}\frac{\partial T}{\partial\phi}-\frac{\sin\theta}{R}\frac{\partial S}{\partial\mu}\right)+\hat{\phi}\left(\sin\theta\frac{\partial T}{\partial\mu}+\frac{1}{R\sin\theta}\frac{\partial S}{\partial\phi}\right),
\end{equation}
and $\boldsymbol{\nabla}\cdot\mathbf{\tilde{B}}=0$. With the above
assumptions, the non-axisymmetric part of the vector-potential reads, 

\begin{eqnarray}
\tilde{\mathbf{A}} & = & \hat{r}T+\boldsymbol{\nabla}\times\left(\hat{r}S\right)\\
 & = & \hat{r}T+\frac{\hat{\theta}}{R\sin\theta}\frac{\partial S}{\partial\phi}+\hat{\phi}\frac{\sin\theta}{R}\frac{\partial S}{\partial\mu}.\nonumber 
\end{eqnarray}
 {Analysis of the solar observations (e.g., \citealt{Pipin2019a})
shows that the ${\displaystyle \frac{\partial rS}{\partial r}}$ term
is related to the non-potential component of the nonaxisymmetric magnetic
field of the solar active regions. Therefore the radial derivative
of $S$ can provide the essential contribution to the surface magnetic
field geometry and energetic of the surface magnetic activity. In
principle, we can neglect the uniform part of $S$ for the sake of
its derivative. However, in this case we would lose the radial component
of the nonaxisymmetric magnetic field. The solar observations show
that the radial magnetic field in solar active regions is dominant
\citep{Scherrer2012}. The model reflect this basic property but it
fails to unveil the radial structure of the nonaxisymmetric magnetic
field.}

\subsection{Dynamo equations}

 {With our representation of the mean electromotive force in
form of the Eq(\ref{eq:simpE}), the full version of the dynamo equations
for the axisymmetric magnetic field reads as follows,
\begin{eqnarray}
\partial_{t}A & = & \overline{\alpha\left\langle B_{\phi}\right\rangle }+\overline{\alpha_{\beta}\left\langle B_{\phi}\right\rangle }+\eta_{T}\Delta'A+\frac{V_{\beta}}{r}\frac{\partial\left(rA\right)}{\partial r}+\overline{V_{\beta}\tilde{B}_{\theta}},\label{eq:A}\\
\partial_{t}B & = & \frac{\sin\theta}{r}\frac{\partial\left(r\sin\theta A,\Omega\right)}{\partial\left(r,\mu\right)}+\eta_{T}\Delta'B-\frac{1}{r}\frac{\partial}{\partial r}r^{2}\left(V_{\beta}B+\overline{V_{\beta}\tilde{B}_{\phi}}\right)\label{eq:B}\\
 & + & \frac{1}{r}\frac{\partial}{\partial r}r\overline{\alpha\left\langle B_{\theta}\right\rangle }+\frac{\sin\theta}{r}\frac{\partial}{\partial\mu}\overline{\alpha\left\langle B_{r}\right\rangle }\nonumber 
\end{eqnarray}
where, $\Delta'=\Delta-{\displaystyle \frac{1}{r^{2}\sin^{2}\theta}}$.
For the sake of brevity, all the $\alpha$-effect and magnetic buoyancy
terms are written explicitly via the magnetic field components. Besides,
these contributions contains effects of the nonlinear coupling between
the axisymmetric and nonaxisymmetric modes of magnetic field. For
example, we have $\overline{\alpha\left\langle B\right\rangle _{\phi}}=\overline{\alpha}B+\overline{\tilde{\alpha}\tilde{B}_{\phi}}$
and the same is for other similar terms. Note that the second part
in this formula, the term $\overline{\tilde{\alpha}\tilde{B}_{\phi}}$
, as well the terms like $\overline{\alpha_{\beta}\left\langle B\right\rangle _{\phi}}$
and similar ones, which are related to the magnetic buoyancy, are
usually ignored in standard mean-field dynamo models. In general,
we can see some similarity of the effect due to $\overline{\tilde{\alpha}\tilde{B}_{\phi}}$
and the non-local $\alpha$-effect employed in the flux-transport
models \citep{Cameron17}. In our approach, we explicitly take into
account dynamics of the nonaxisymmetric magnetic field and their averaged
effect on the evolution of the large-scale magnetic field. }

 {Applying our simplifications to the Eqs(\ref{eq:A} and \ref{eq:B})
we arrive to the ``no-r'' equations for evolution of the axisymmetric
magnetic field:
\begin{eqnarray}
\partial_{t}A & = & \alpha\mu\left\langle B_{\phi}\right\rangle +\eta_{T}\frac{\sin^{2}\theta}{R^{2}}\frac{\partial^{2}\left(\sin\theta A\right)}{\partial\mu^{2}}-\frac{V_{\beta}}{R}A-\frac{A}{\tau},\label{eq:at}
\end{eqnarray}
}

 {
\begin{eqnarray}
\partial_{t}B & =- & \sin\theta\frac{\partial\Omega}{\partial r}\frac{\partial\left(\sin\theta A\right)}{\partial\mu}+\eta_{T}\frac{\sin^{2}\theta}{R^{2}}\frac{\partial^{2}\left(\sin\theta B\right)}{\partial\mu^{2}}\label{eq:bt}\\
 &  & +\frac{\sin\theta}{R}\frac{\partial}{\partial\mu}\overline{\alpha\left\langle B_{r}\right\rangle }+\frac{\overline{\alpha\left\langle B_{\theta}\right\rangle }}{R}-\frac{1}{R}V_{\beta}\left\langle B_{\phi}\right\rangle -\frac{B}{\tau}\nonumber 
\end{eqnarray}
On the one hand, in the shortened version of the dynamo equations
we keep the dominant dynamo effect caused by the radial gradient of
the angular velocity. The $\tau$-terms in Eqs(\ref{eq:bt},\ref{eq:at})
were suggested by \citet{Moss2008} to account for turbulent diffusion
in radial direction. Similarly to the cited paper we put ${\displaystyle \tau=3\frac{R^{2}}{\eta_{T}}}$.
The magnetic buoyancy effect results to the physically similar terms
(see, \citealt{Noyes1984}). On the other hand, we assume that the
magnetic field and $\alpha$-effect are uniform in the radial direction.
This is an inconsistency of the ``no-r'' approach. }

 {To get the evolution equation for the nonaxisymmetric potential
$S$ we substitute the Eq(\ref{eq:bdec}) into Eq(\ref{eq:mfe}),
and then calculate the scalar product with vector $\mathbf{\hat{r}}$.
Similarly, equation for $T$ is obtained by taking curl of Eq(\ref{eq:mfe})
and then the scalar product with vector $\mathbf{\hat{r}}$. The procedure
is described in detail by \citet{Krause1980}, also, see, \citet{Bigazzi2004}
and \citet{Pipin2015a}. The equations for the potentials T and S
are }

 {
\begin{eqnarray}
\partial_{t}\Delta_{\Omega}S & = & \Delta_{\Omega}U^{(U)}+\Delta_{\Omega}U^{(\mathcal{E})},\label{eq:S}\\
\partial_{t}\Delta_{\Omega}T & = & \Delta_{\Omega}V^{(U)}+\Delta_{\Omega}V^{(\mathcal{E})},\label{eq:T}
\end{eqnarray}
where we introduce the new notations 
\begin{eqnarray}
\Delta_{\Omega}V^{(U)} & = & -\hat{\mathbf{r}}\cdot\boldsymbol{\nabla}\times\boldsymbol{\nabla}\times\left(\mathbf{\overline{U}}\times\mathbf{\tilde{\mathbf{B}}}\right),\label{eq:vu}\\
\Delta_{\Omega}V^{(\mathcal{E})} & = & -\hat{\mathbf{r}}\cdot\boldsymbol{\nabla}\times\boldsymbol{\nabla}\times\boldsymbol{\boldsymbol{\mathcal{E}}},\label{eq:ve}\\
\Delta_{\Omega}U^{(U)} & = & -\hat{\mathbf{r}}\cdot\boldsymbol{\nabla}\times\left(\mathbf{\overline{U}}\times\tilde{\mathbf{B}}\right),\label{eq:uu}\\
\Delta_{\Omega}U^{(\mathcal{E})} & = & -\hat{\mathbf{r}}\cdot\boldsymbol{\nabla}\times\boldsymbol{\boldsymbol{\mathcal{E}}}.\label{eq:ue}
\end{eqnarray}
The effect of the differential rotation on the nonaxisymmetric magnetic
field reads as follows,
\begin{eqnarray*}
\Delta_{\Omega}V^{(U)} & = & -\Delta_{\Omega}\Omega\frac{\partial T}{\partial\phi}+\Delta_{\Omega}\left\{ \left(\boldsymbol{\Omega}\cdot\boldsymbol{\nabla}\right)F_{S}-\left(\mathbf{\hat{r}}\cdot\boldsymbol{\Omega}\right)\Delta S\right\} \\
 &  & -\boldsymbol{\nabla}\cdot\left\{ \frac{\boldsymbol{\Omega}}{\Omega}\left(\mathbf{\hat{r}\cdot\boldsymbol{\nabla}}\right)-\left(\mathbf{\hat{r}}\cdot\frac{\boldsymbol{\Omega}}{\Omega}\right)\boldsymbol{\nabla}\right\} \left(\Omega\Delta_{\Omega}S\right)\\
\Delta_{\Omega}U^{(U)} & = & -\frac{\partial}{\partial\phi}\left(\Omega\Delta_{\Omega}S\right)
\end{eqnarray*}
For the turbulent magnetic diffusivity, magnetic buoyancy and $\alpha$
- effect we get
\begin{eqnarray}
\Delta_{\Omega}V^{(\mathcal{E})} & = & -\Delta_{\Omega}\left(\frac{1}{r}\frac{\partial}{\partial r}\left(\eta_{T}\frac{\partial}{\partial r}rT\right)+\frac{\eta_{T}}{r^{2}}\Delta_{\Omega}T\right)\label{eq:VE}\\
 & - & \frac{1}{r\sin\theta}\frac{\partial}{\partial\phi}\frac{\partial}{\partial r}\left(r\left\langle B_{\theta}\right\rangle V_{\beta}\right)-\frac{\partial}{\partial\mu}\left(\frac{\sin\theta}{r}\frac{\partial}{\partial r}\left(r\left\langle B_{\phi}\right\rangle V_{\beta}\right)\right)\nonumber \\
 & + & \frac{1}{r}\Delta_{\Omega}\alpha\left\langle B_{r}\right\rangle +\frac{1}{r}\frac{\partial}{\partial r}r\frac{\partial}{\partial\mu}\alpha\left\langle B_{\theta}\right\rangle -\frac{1}{r}\frac{\partial}{\partial r}r\left[\frac{1}{\sin\theta}\frac{\partial}{\partial\phi}\left(\alpha+\alpha_{\beta}\right)\left\langle B_{\phi}\right\rangle \right],\nonumber 
\end{eqnarray}
\begin{eqnarray}
\Delta_{\Omega}U^{(\mathcal{E})} & = & \eta_{T}\Delta_{\Omega}\Delta S-\frac{1}{\sin\theta}\frac{\partial}{\partial\phi}\left(\left\langle B_{\phi}\right\rangle V_{\beta}\right)+\frac{\partial}{\partial\mu}\left(\sin\theta\left\langle B_{\theta}\right\rangle V_{\beta}\right)\label{eq:UE}\\
 & + & \frac{\partial}{\partial\mu}\left(\alpha+\alpha_{\beta}\right)\left\langle B_{\phi}\right\rangle +\frac{1}{\sin\theta}\frac{\partial}{\partial\phi}\alpha\left\langle B_{\theta}\right\rangle .\nonumber 
\end{eqnarray}
Applying our ``no-r'' approach to the Eqs(\ref{eq:S} and \ref{eq:T})
we get}

 {
\begin{eqnarray}
\partial_{t}\Delta_{\Omega}T & = & -\Delta_{\Omega}\Omega\frac{\partial T}{\partial\phi}+\frac{\eta_{T}}{R^{2}}\Delta_{\Omega}^{2}T-\frac{\Delta_{\Omega}T}{\tau}\label{eq:Tt}\\
 & - & \frac{1}{R}\frac{\partial\Omega}{\partial r}\sin^{2}\theta\frac{\partial\Delta_{\Omega}S}{\partial\mu}-\frac{1}{R}\frac{\partial}{\partial\phi}\left[\frac{\alpha+\alpha_{\beta}}{\sin\theta}\left\langle B_{\phi}\right\rangle \right]\nonumber \\
 & + & \Delta_{\Omega}\frac{\alpha\left\langle B_{r}\right\rangle }{R}+\frac{1}{R}\frac{\partial}{\partial\mu}\alpha\left\langle B_{\theta}\right\rangle \nonumber \\
 & - & \frac{1}{R\sin\theta}\frac{\partial}{\partial\phi}\left\langle B_{\theta}\right\rangle V_{\beta}-\frac{1}{R}\frac{\partial}{\partial\mu}\left(\sin\theta\left\langle B_{\phi}\right\rangle V_{\beta}\right),\nonumber 
\end{eqnarray}
}

 {
\begin{eqnarray}
\partial_{t}\Delta_{\Omega}S & = & -\left(\Omega\Delta_{\Omega}\frac{\partial}{\partial\phi}S\right)+\frac{\eta_{T}}{R^{2}}\Delta_{\Omega}^{2}S-\frac{\Delta_{\Omega}S}{\tau}\label{eq:St}\\
 &  & +\frac{\partial}{\partial\mu}\left(\alpha+\alpha_{\beta}\right)\left\langle B_{\phi}\right\rangle +\frac{\partial}{\partial\phi}\frac{\alpha\left\langle B_{\theta}\right\rangle }{\sin\theta}\nonumber \\
 & - & \frac{1}{\sin\theta}\frac{\partial}{\partial\phi}\left(\left\langle B_{\phi}\right\rangle V_{\beta}\right)+\frac{\partial}{\partial\mu}\left(\sin\theta\left\langle B_{\theta}\right\rangle V_{\beta}\right).\nonumber 
\end{eqnarray}
Here, same as before, we write all the $\alpha$'s and magnetic buoyancy
terms explicitly via the magnetic field components. The numerical
solution of the dynamo problem treats these terms in the same way.
Our above comments about the shortened version of the axisymmetric
dynamo should be applied to the Eqs(\ref{eq:Tt} and \ref{eq:St}),
as well. In addition, to simulate stretching of nonaxisymmetric magnetic
field by the surface differential rotation we take into account the
latitudinal dependence of the angular velocity $\Omega=1-0.25\sin^{2}\theta\Omega$. }

\subsection{Helicity evolution equation\label{helint}}

We start from the mean-field evolution equations for the large-scale
magnetic field:

\begin{eqnarray}
\partial_{t}\left\langle \mathbf{B}\right\rangle  & = & \mathbf{\nabla}\times\left(\mathbf{\mathbf{\mathbf{\mathcal{E}}}+}\left\langle \mathbf{U}\right\rangle \times\left\langle \mathbf{B}\right\rangle -\eta\mathbf{\nabla}\times\left\langle \mathbf{B}\right\rangle \right)\label{eq:mfe-1}
\end{eqnarray}
For the time derivative of the magnetic helicity helicity of the large-scale
field in a volume we get,
\begin{eqnarray}
\frac{d}{dt}\int\left\langle \mathbf{A}\right\rangle \cdot\left\langle \mathbf{B}\right\rangle \mathrm{dV} & = & \int\left(\left\langle \mathbf{A}\right\rangle \cdot\partial_{t}\left\langle \mathbf{B}\right\rangle +\left\langle \mathbf{B}\right\rangle \cdot\partial_{t}\left\langle \mathbf{A}\right\rangle +\mathbf{\nabla\cdot}\left\langle \mathbf{U}\right\rangle \left(\left\langle \mathbf{A}\right\rangle \cdot\left\langle \mathbf{B}\right\rangle \right)\right)\mathrm{dV}\label{eq:int}\\
 & = & \int\left\{ 2\left\langle \mathbf{A}\right\rangle \cdot\partial_{t}\left\langle \mathbf{B}\right\rangle +\mathbf{\nabla\cdot}\left\langle \mathbf{U}\right\rangle \left(\left\langle \mathbf{A}\right\rangle \cdot\left\langle \mathbf{B}\right\rangle \right)\right\} dV+\oint\left(\left\langle \mathbf{A}\right\rangle \times\partial_{t}\left\langle \mathbf{A}\right\rangle \right)\cdot\mathbf{n}\mathrm{dS}\nonumber 
\end{eqnarray}
 Using the mean-field evolution equation we obtain:
\begin{eqnarray}
\frac{d}{dt}\int\left\langle \mathbf{A}\right\rangle \cdot\left\langle \mathbf{B}\right\rangle \mathrm{dV} & = & 2\int\left(\mathbf{\mathbf{\mathcal{E}}}\cdot\left\langle \mathbf{B}\right\rangle -\eta\left\langle \mathbf{J}\right\rangle \cdot\left\langle \mathbf{B}\right\rangle \right)\mathrm{dV}\label{eq:int2}\\
 & + & 2\int\mathbf{\nabla\cdot}\left(\mathbf{\mathbf{\mathcal{E}}}\times\left\langle \mathbf{A}\right\rangle +\left\langle \mathbf{B}\right\rangle \left(\left\langle \mathbf{A}\right\rangle \cdot\left\langle \mathbf{U}\right\rangle \right)-\eta\left\langle \mathbf{J}\right\rangle \times\left\langle \mathbf{A}\right\rangle \right)\mathrm{dV}\nonumber \\
 & - & \oint\left(\left\langle \mathbf{A}\right\rangle \cdot\left\langle \mathbf{B}\right\rangle \right)\left(\left\langle \mathbf{U}\right\rangle \cdot\mathbf{n}\right)\mathrm{dS}+\oint\left(\left\langle \mathbf{A}\right\rangle \times\partial_{t}\left\langle \mathbf{A}\right\rangle \right)\cdot\mathbf{n}\mathrm{dS},\nonumber 
\end{eqnarray}
where we denote $\left\langle \mathbf{J}\right\rangle =\mathbf{\nabla}\times\left\langle \mathbf{B}\right\rangle $.
 {Interestingly, that the diffusive part of the flux $\mathbf{\mathbf{\mathcal{E}}}\times\left\langle \mathbf{A}\right\rangle =-\eta_{T}\left\langle \mathbf{J}\right\rangle \times\left\langle \mathbf{A}\right\rangle $
contains the flux of large-scale magnetic helicity which can be represented
though the gradient of the large-scale helicity. Indeed, 
\begin{eqnarray}
\left\langle \mathbf{A}\right\rangle \times\mathbf{\nabla}\times\left\langle \mathbf{B}\right\rangle  & = & \left(\left\langle \mathbf{A}\right\rangle _{i}\mathbf{\nabla}\left\langle \mathbf{B}\right\rangle _{i}\right)-\left(\left\langle \mathbf{A}\right\rangle \cdot\mathbf{\nabla}\right)\left\langle \mathbf{B}\right\rangle ,\label{eq:AxB}\\
 & = & \mathbf{\nabla}\left(\left\langle \mathbf{A}\right\rangle \cdot\left\langle \mathbf{B}\right\rangle \right)-\left(\left\langle \mathbf{B}\right\rangle _{i}\mathbf{\nabla}\left\langle \mathbf{A}\right\rangle _{i}\right)-\left(\left\langle \mathbf{A}\right\rangle \cdot\mathbf{\nabla}\right)\left\langle \mathbf{B}\right\rangle ,\nonumber 
\end{eqnarray}
therefore we have 
\begin{equation}
\mathbf{\nabla\cdot}\left(\mathbf{\mathbf{-\eta_{T}\left\langle \mathbf{J}\right\rangle \times\left\langle \mathbf{A}\right\rangle }}\right)=\mathbf{\nabla\cdot}\left(\eta_{T}\mathbf{\nabla}\left(\left\langle \mathbf{A}\right\rangle \cdot\left\langle \mathbf{B}\right\rangle \right)-\eta_{T}\left(\left\langle \mathbf{B}\right\rangle _{i}\mathbf{\nabla}\left\langle \mathbf{A}\right\rangle _{i}+\left(\left\langle \mathbf{A}\right\rangle \cdot\mathbf{\nabla}\right)\left\langle \mathbf{B}\right\rangle \right)\right).\label{eq:AxB1}
\end{equation}
 The last equation shows that the contribution of the gradient flux
of the large-scale helicity is a part of a more general expression.
In our paper we do not concern the meaning of this flux for the dynamo
model. The helicity flux from $-\eta_{T}\left\langle \mathbf{J}\right\rangle \times\left\langle \mathbf{A}\right\rangle $
was analyzed recently by \citet{Hawkes2019AA} using the surface flux-transport
model. In our model there is some arbitrariness about contribution
of the gradient flux of the large-scale helicity density in evolution
of the total magnetic helicity. Our paper employ} the heuristic integral
conservation law for the total helicity, $\left\langle \chi\right\rangle ^{(tot)}=\left\langle \chi\right\rangle +\left\langle \mathbf{A}\right\rangle \cdot\left\langle \mathbf{B}\right\rangle $,
evolution originally suggested by \citet{Hubbard2012}, 
\begin{equation}
\frac{d}{dt}\int\left(\left\langle \chi\right\rangle +\left\langle \mathbf{A}\right\rangle \cdot\left\langle \mathbf{B}\right\rangle \right)\mathrm{dV}=-\frac{1}{R_{m}\tau_{c}}\int\left\langle \chi\right\rangle \mathrm{dV}-2\eta\int\left\langle \mathbf{B}\right\rangle \cdot\left\langle \mathbf{J}\right\rangle \mathrm{dV}-\int\mathbf{\nabla\cdot}\mathbf{\mathbf{\mathcal{F}}}^{\chi}\mathrm{dV},\label{eq:int3}
\end{equation}
where  {the heuristic term,} $\mathbf{\mathbf{\mathcal{F}}}^{\chi}=-\eta_{\chi}\mathbf{\nabla}\left\langle \chi\right\rangle ^{(tot)}$
was suggested in the above cited paper in substitution of the third
order correlations which involve products of the small-scale turbulent
parts of the vector potential, magnetic field and currents (also,
see, \citealt{Kleeorin1999}). Subtracting, the Eq(\ref{eq:int2})
from Eq(\ref{eq:int3}) we get,
\begin{eqnarray}
\left(\frac{d}{dt}+\frac{1}{R_{m}\tau_{c}}\right)\int\left\langle \chi\right\rangle \mathrm{dV} & = & -2\int\mathbf{\mathbf{\mathbf{\mathcal{E}}}}\cdot\left\langle \mathbf{B}\right\rangle \mathrm{dV}-\int\boldsymbol{\nabla\cdot}\boldsymbol{\boldsymbol{\mathcal{F}}}^{\chi}\mathrm{dV}\label{eq:int-cons}\\
 & - & 2\oint\left(\mathbf{\mathbf{\mathcal{E}}}\times\mathbf{\left\langle A\right\rangle }\right)\cdot\mathbf{n}\mathrm{dS}-2\oint\left(\left\langle \mathbf{A}\right\rangle \cdot\left\langle \mathbf{U}\right\rangle \right)\left(\left\langle \mathbf{B}\right\rangle \cdot\mathbf{n}\right)\mathrm{dS}\nonumber \\
 & - & 2\eta\oint\left(\left\langle \mathbf{A}\right\rangle \times\left\langle \mathbf{J}\right\rangle \right)\cdot\mathbf{n}\mathrm{dS}+\oint\left(\left\langle \mathbf{A}\right\rangle \cdot\left\langle \mathbf{B}\right\rangle \right)\left(\left\langle \mathbf{U}\right\rangle \cdot\mathbf{n}\right)\mathrm{dS}\nonumber \\
 & - & \oint\left(\left\langle \mathbf{A}\right\rangle \times\partial_{t}\left\langle \mathbf{A}\right\rangle \right)\cdot\mathbf{n}\mathrm{dS},
\end{eqnarray}
The last term, in the above equation, gets zero for a divergency free
on the surface vector potentials \citep{Berger2018}. In our definitions,
the vector potential is consisted of a sum of the divergency-free
components and the pure radial components (see, the Eqs.(\ref{eq:app},\ref{eq:bdec})).
Therefore the last term in the equation \ref{eq:int-cons} is zero,
as well.
\end{document}